\documentclass[%
preprint,
frontmatter,
 amsmath,amssymb,3p,
12pt]{elsarticle}

\usepackage{amssymb}
\usepackage{graphicx}
\usepackage{dcolumn}
\usepackage{comment}
\usepackage{color}
\usepackage{float}
\usepackage[utf8]{inputenc}
\usepackage{graphicx}
\usepackage{subcaption}
\usepackage{multirow}
\usepackage{bibunits}
\usepackage{bigints}
\graphicspath{{Images_new/}}
\usepackage{placeins}
\usepackage{afterpage}

\usepackage[pdftex,breaklinks,colorlinks,
citecolor=blue,
urlcolor=blue,
pdfsubject={LaTeX}]{hyperref}

\begin{document}

\begin{frontmatter}
\title{Vacancy diffusion in multi-principal element alloys: the role of chemical disorder in the ordered lattice}

\author[1]{Spencer L. Thomas\corref{cor1}}
\ead{slthom23@ncsu.edu}
\address[1]{Department of Materials Science and Engineering \\
 North Carolina State University}
 
\author[1]{Srikanth Patala\corref{cor1}}%
\ead{spatala@ncsu.edu}
\cortext[cor1]{Corresponding authors}
\begin{abstract}

Many of the purported virtues of Multi-Principal Element Alloys (MPEAs), such as corrosion, high-temperature oxidation and irradiation resistance, are highly sensitive to vacancy diffusivity. Similarly, solute interdiffusion is governed by vacancy diffusion – it is often unclear whether MPEAs are truly stable, or effectively stabilized by slow interdiffusion. The considerable composition space afforded to these alloys makes optimizing for desired properties a daunting task; theoretical and computational tools are necessary to guide alloy development. For diffusion, such tools depend on both a knowledge of the vacancy migration barriers within a given alloy and an understanding of how these barriers influence vacancy diffusivity. We present a generalized theory of vacancy diffusion in rugged energy landscapes, paired with Kinetic Monte Carlo simulations of MPEA vacancy diffusion. The barrier energy statistics are informed by nudged elastic band calculations in the equiatomic CoNiCrFeMn alloy. Theory and simulations show that vacancy diffusion in solid-solution MPEAs is not necessarily sluggish, but can potentially be tuned, and that trap models are an insufficient explanation for sluggish diffusion in the CoNiCrFeMn MPEA. These results also show that any model that endeavors to faithfully represent diffusion-related phenomena must account for the full nature of the energy landscape, not just the migration barriers.

\begin{description}
\item[Keywords] Mutli-Principal Element Alloys,  High-Entropy Alloys, Diffusion, Random Walk, Correlation, Disordered Materials

\end{description}
\end{abstract}
\end{frontmatter}

\newpage
\clearpage

\section{Introduction}
\label{sec:Intro}

In recent years, there has been considerable interest in designing novel
structural alloys by combining multiple elements in high concentrations,
resulting in the so-called multi-principal element alloys (MPEAs) (also referred
to as high-entropy alloys or complex concentrated alloys). These alloys have been shown to exhibit
exceptional properties, including novel mechanical properties
\cite{senkov2011mechanical, gali2013tensile}, fracture toughness
\cite{gludovatz2014fracture, zhang2015nanoscale}, creep resistance
\cite{lee2016spherical, ma2016nanoindentation, zhang2016nanoindentation,
  cao2016influence}, radiation damage resistance \cite{egami2014irradiation,
  xia2015irradiation1, xia2015irradiation, granberg2016mechanism,
  kumar2016microstructural}, and oxidation resistance
\cite{holcomb2015oxidation, kai2016oxidation, laplanche2016oxidation}.
The property enhancements in MPEAs, as well as their thermal stability, are
usually attributed to four core properties: (a) high mixing entropy, (b) lattice
distortions, (c) sluggish diffusion, and (d) cocktail effects
\cite{miracle2017high}. Among these effects, sluggish diffusion is poorly
understood and may be the most controversial in the MPEA community. It was long assumed
that slow diffusion was an inevitable consequence of traps and obstacles that arise from the
disordered lattice \cite{tsai2013sluggish}; disordered (or rough) energy landscapes have indeed proven capable of inducing sluggish kinetics
\cite{ansari1985protein,haus1987diffusion,zwanzig1988diffusion}, but recent
experimental data has suggested that not all MPEAs exhibit slow diffusion
\cite{dkabrowa2019demystifying, miracle2019high}.

Quantifying atomic diffusion in MPEAs is important, as the solid-solution
phase with the highest mixing entropy does not always have the lowest Gibbs free
energy \cite{zhang2014understanding} and complex phases may precipitate in MPEAs
after annealing treatments. For example, Cheng et al. \cite{cheng2016high,
  cheng2016high1} attribute the remarkable thermal stability of the amorphous
structure in Ge$_{x}$TiZrNbTa and BTiZrNbTa thin films to the combination of
high entropy, significant atomic size differences, and sluggish
diffusion. In \cite{zhao2018thermal}, Zhao et al. studied the coarsening of
L1$_2$ precipitates in a face-centered-cubic
{(NiCoFeCr)}{$_{94}$}{Ti}{$_2$}{Al}{$_4$} MPEA at temperatures ranging between 750$^{\circ}$ and
825$^{\circ}$C. They concluded that, owing to sluggish diffusion, L1$_2$ precipitate coarsening was much slower in the MPEA than in conventional Ni-based alloys. 


Beyond thermal stability, understanding the role of structural disorder on diffusivity is also important for the performance of MPEAs. 
For example, high-entropy alloy nitrides have been proposed to be effective diffusion barrier
coatings \cite{kumar2020diffusion}; the reduced diffusivity of these materials is attributed to lattice distortions caused by
multiple principal elements. MPEAs have also been proposed as replacements for
conventional alloys in high temperature applications \cite{praveen2018high} and in 
advanced nuclear reactors \cite{lu2019promising, barr2018exploring}. Under
extreme environments, the diffusivity of point-defects (vacancies) governs the
performance and the microstructural stability of MPEAs. Therefore, determination of the kinetics of phase transformations and operating temperatures for these alloys demands a thorough understanding of the transport coefficients \cite{jeong2019high}.

In this article, we investigate the diffusivity of vacancies in model MPEAs.
More often than not, diffusion in crystalline solids is mediated by vacancy
point defects. Therefore, the kinetics of phase transformations and
microstructural evolution will be greatly influenced by vacancy diffusivity  \cite{balluffi2005kinetics} in MPEAs.
From a theoretical standpoint, as far as the authors are aware of, the models
for calculating diffusion coefficients in MPEAs are primarily based on the
formulation of linear irreversible thermodynamics and Onsager coefficients
\cite{moleko1989self, allnatt2016high, trinkle2018variational}.
These frameworks are developed for multi-component alloys and
  are purported to be valid over a wide range of compositions. However, these
  models make the underlying assumption that there is a constant exchange rate
  (i.e a constant migration barrier) between any two diffusing species
  present in the alloys. The theoretical results are then validated using
  Kinetic Monte Carlo simulations with the same underlying assumptions, i.e. the
  exchange rates are fixed and constant. However, it is also true that the randomness in the energy landscape plays a
crucial role in influencing transport coefficients as evidenced in solid-state
and biological systems \cite{summerfield1981effective,odagaki1981coherent,
  webman1981effective, banerjee2014diffusion, seki2015relationship,
  seki2016anomalous}. In MPEAs, the migration barrier energies for vacancies (or
self-interstitials) are randomly distributed (e.g. as shown in Figure 3 of Ref.
\cite{osetsky2016specific}). The nature of this distribution will play an
important role in determining the diffusion coefficients in MPEAs.


 The objective of the present article is to provide a random-walk-type \emph{analytical
  results} for understanding vacancy diffusivity in the presence of chemical
disorder on the ordered lattice of MPEAs. Random-walk approaches provide the
atomistic detail necessary to account for the disorder (or roughness) present in the 
migration barrier energies. In Section~\ref{sec:background}, we present some
background on diffusion models in the presence of disorder (roughness) in energy
landscapes. In Section \ref{sec:neb}, we provide the statistics of vacancy migration
barrier energies for a model MPEA that will motivate the KMC simulations
(Section~\ref{sec:kmc}) and theoretical analysis (Section~\ref{sec:theory}). The theoretical
results presented in this article are anticipated to provide the necessary
foundation for understanding diffusion in multi-component
solid solutions.


\section{Background}
\label{sec:background}

The analytical investigation of random energy landscapes and 
transport coefficients originates in Robert Zwanzig's work on protein
dynamics\cite{zwanzig1988diffusion}. Zwanzig showed that the diffusivity in a
randomly rough potential $U$ is $D = D_{\circ} \exp \left( -(\epsilon/k_B T)^2 \right)$, where $D_{\circ}$ is the diffusion coefficient in a landscape with uniform barriers and $\epsilon^2 = \langle U^2 \rangle$. Zwanzig's theory suggests an exponential
reduction in the diffusion coefficient with the (squared)-amplitude of
roughness. While this result is elegant and simple, Zwanzig also invoked
certain assumptions (e.g. local averaging of the rough landscape) that 
may not always be valid. Additionally, these assumptions
break down when correlations in the random-walks are considered.

\begin{figure*}[h!]
    \centering
    \includegraphics[width=1.0\linewidth]{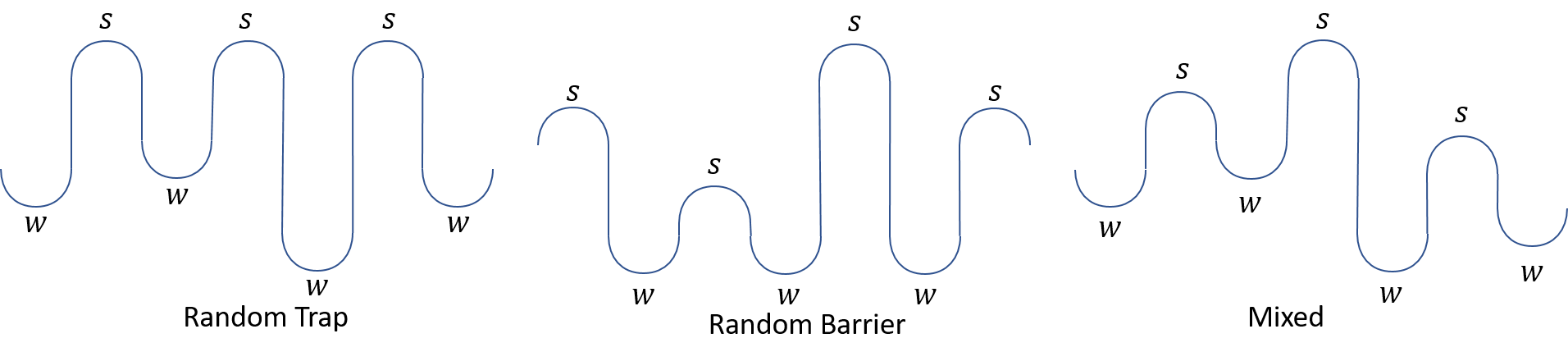}
    \caption{Energy landscapes with (a) random traps, (b) random barriers, and (c) the general case.
    \label{fig:rtb}}
\end{figure*}

Analyses of random walks in disordered lattices has also been of considerable
interest with applications in solid-state physics
\cite{summerfield1981effective,odagaki1981coherent, webman1981effective},
biology \cite{banerjee2014diffusion, seki2015relationship, seki2016anomalous},
and in amorphous materials \cite{mussawisade1997combination}. These
investigations use the Effective Medium Approximation (EMA) to solve 
for a mean transition rate that satisfies the self-consistency 
condition \cite{webman1981effective,haus1987diffusion} for a given energy landscape. 
These landscapes are often broken into two categories -- 
random-trap models, and random-barrier models (see Fig.~\ref{fig:rtb}).

Random-trap (RT) models refer to energy landscapes where the transition-state (saddle-point) energies $s$ are fixed to a constant value and the site-energies $w$ are assumed to follow a Gaussian distribution (see Fig.~\ref{fig:rtb}(a)) with a given standard deviation $\sigma_w$. 
Direct integration (comparing the average transition rate to a uniform distribition of barriers with the same mean) yields $D/D_\circ = \exp(-x_w)$, where $x_w = (\sigma_w/k_B T)^2$, mirroring Zwanzig's model. Intuitively, the reduction in diffusivity can be explained as the trapping of the diffusing species in low energy sites. A defining characteristic of RT models is that, from a given state, \textbf{every transition out of that state has the same barrier $E$}.

Random-barrier (RB) models correspond to energy landscapes where the site-energies $w$ have the same fixed value and the transition-state energies $s$ are randomly distributed (see Fig.~\ref{fig:rtb}(b)). For example, effective medium solutions were proposed for electron transport, employing landscapes with a uniform distribution of barriers from 0 to some critical energy $E_c$ \cite{ambaye1995asymptotic}. In this study, it was shown that a wider distribution of barriers yields a smaller diffusion constant, but this is obvious as the mean barrier is also increasing. 
Mussawsade et al. applied the EMA to a Gaussian-distributed random barrier model \cite{mussawisade1997combination}. 
Here, the mean barrier was chosen for a given distribution such that $95\%$ of the distribution was positive (as negative barriers are unphysical). 
Naturally, they also showed slower diffusion with an increasing distribution width as the mean barrier was also increasing. Symmetry is a defining feature of RB models - If $E_{A\rightarrow B}$ is the energy of a transition from state A to B, then in the RB model, $E_{A\rightarrow B} = E_{B\rightarrow A}$.

It is plausible that some of the initial justifications for sluggish diffusion
in MPEAs were inspired by the analytical results summarized above. The idea of
sluggish diffusion, however, is brought into question recently with many
research groups convinced that sluggish diffusion is not one of the core
properties of MPEAs that fundamentally distinguishes them from conventional
alloys. While this could very well be true, the diffusion of atomic species and
vacancies has been invoked in various recent studies that report property
enhancements in MPEAs. It is necessary, therefore, to determine the
nature of the energy landscape in an MPEA and investigate its role
in influencing transport properties. It is essential that we investigate
these aspects not just for the long-term 
development of MPEAs, but also for the short-term creation 
of computational tools. For example, it is traditional to assume
that vacancy diffusivity can be described by simply computing
the migration energy barriers in the MPEA. However, the present work 
strongly suggests that this is only a partial picture.
\section{Determination of Vacancy Migration Barrier Energies} 
\label{sec:neb}

When a vacancy swaps with an atom, the migration barrier will depend on the species of that atom, as well as those of neighboring atomic sites. The chemical disorder of the alloy then implies a distribution of migration barriers. 
Before an analysis of transport in MPEAs can be conducted, we must actually compute these barriers to vacancy migration in a sample MPEA. We performed 2971 Nudged Elastic Band (NEB) 
\cite{henkelman2000climbing, henkelman2000improved, nakano2008space,maras2016global} calculations of vacancy migration between nearest-neighbor sites, using a MEAM potential for the CoCrFeMnNi High
Entropy Alloy \cite{choi2018understanding} and the LAMMPS atomistic simulation
software \cite{plimpton1995fast}. 

First, we generated a random equiatomic face-centered cubic (FCC) single crystal MPEA consisting of 32000 atoms. We then alternated between performing 100000 monte carlo swaps \cite{sadigh2012scalable} at 1273 K and relaxing the system via conjugate-gradient (CG) energy
minimization \cite{bitzek2006structural,sheppard2008optimization}. This procedure was repeated 500 times to capture any tendency for short-range ordering. After 20 million swaps, there was partial ordering, according to the short-range order (SRO) parameters \cite{fernandez2017short}. The SRO parameters did not change significantly over the next 30 million swaps. 

Once this initial configuration was generated, we iterated through atoms in the system. 
For each iteration, we removed the target atom and relaxed the system, again by CG minimization. In the FCC crystal, each vacancy site has twelve neighbors. 
For each neighbor, a post-migration configuration was created by swapping that neighbor with the vacancy, followed by another relaxation. These served as the initial and final images for the NEB calculations, which were performed with a spring constant of $1$ eV/\AA (we verified this choice with a parametric study; the results were insensitive to the choice of spring constant within a range of $.1$ and $10 $ eV/\AA).

\begin{figure}[h!]
    \centering
    \includegraphics[width=0.6\linewidth]{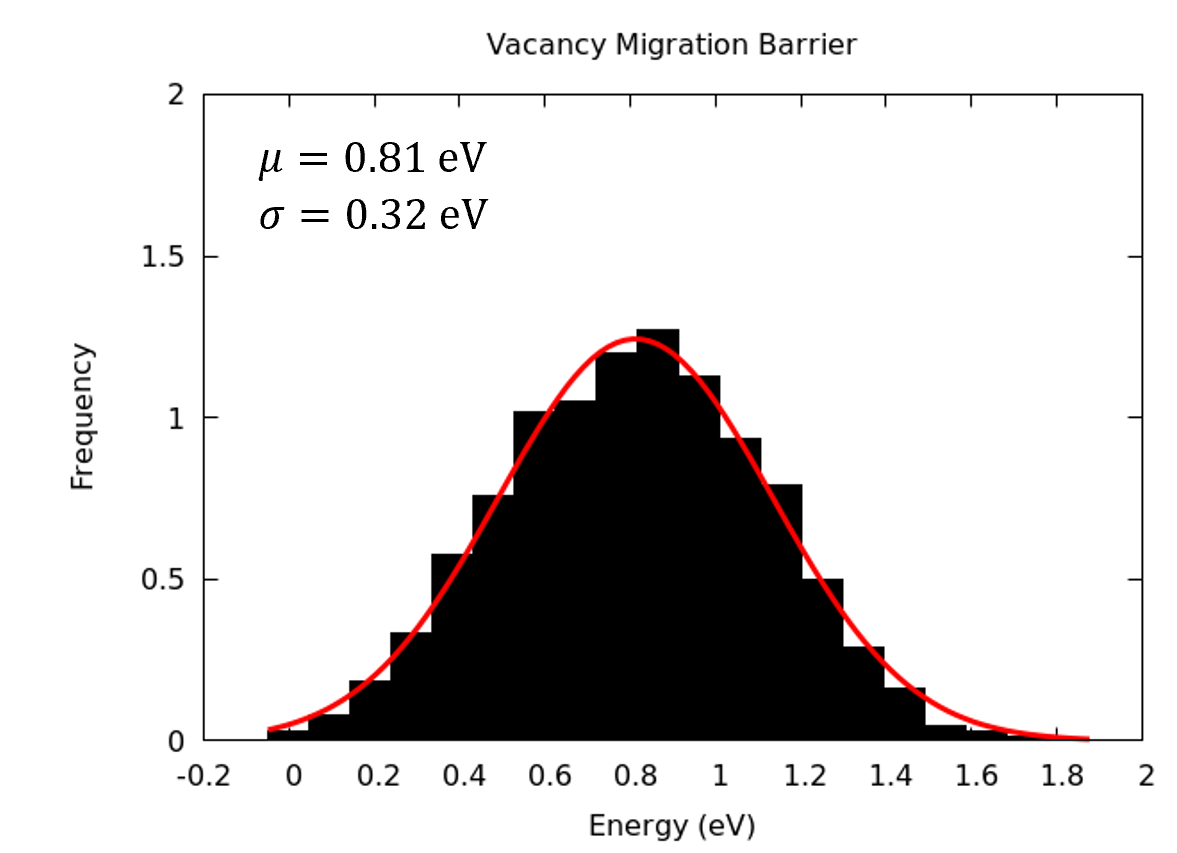}
    \caption{Distribution of vacancy migration barrier energies in
      CoNiFeCrMn MPEA \cite{choi2018understanding}, computed using 2971 Nudged
      Elastic Band (NEB) calculations. The barriers are
      Gaussian distributed; mean and standard deviations, determined from the fit (red) are given.}
    \label{fig:GDist}
\end{figure}

\begin{figure}[h!]
\centering
\begin{tabular}[b]{c}
  \includegraphics[width=.45\linewidth]{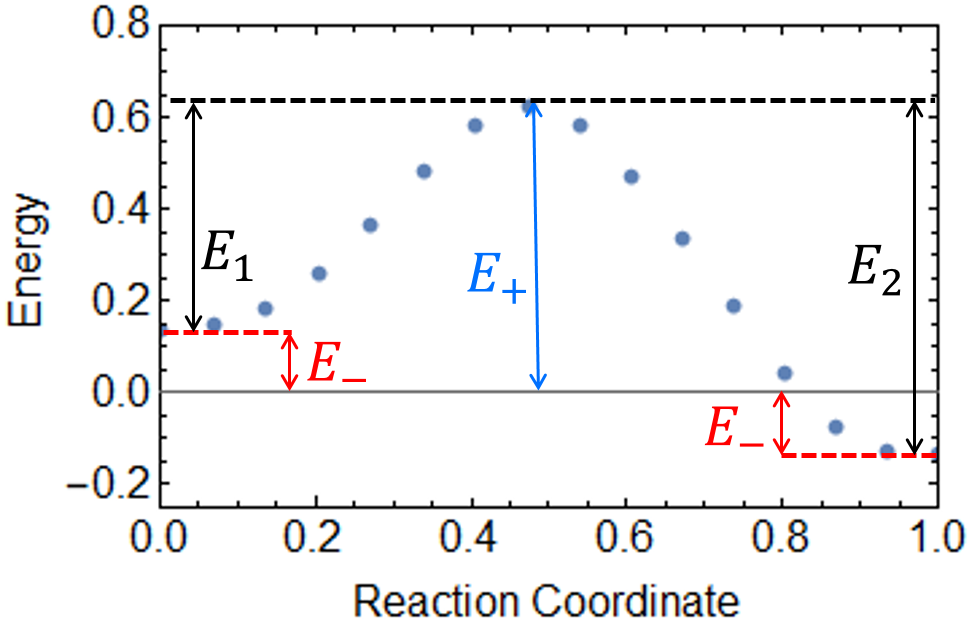} \\
  \small (a)
  \end{tabular} \quad
\begin{tabular}[b]{c}
    \includegraphics[width=.45\linewidth]{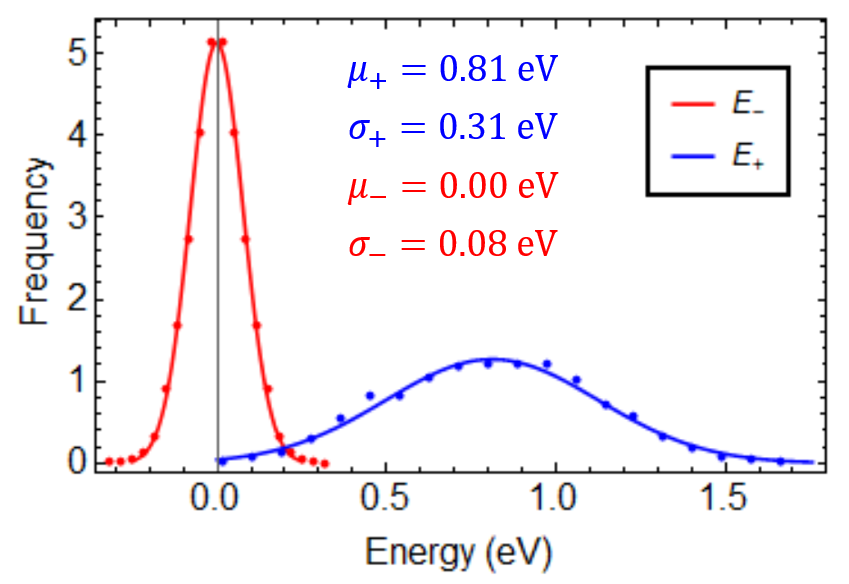} \\
    \small (b)
\end{tabular} 
\quad
\caption{a) Definition of $E_+$ and $E_-$, the symmetric and anti-symmetric barrier components. For a reference energy equal to the mean of the energies of the initial and final state, $s = E_+$ and $w_i = -w_f = E_-$. b) Distribution of
  $E_-$ and $E_+$ for the CoNiFeCrMn HEA \cite{choi2018understanding} computed from 2971 NEB calculations.
}
\label{fig:NEB}
\end{figure}

The vacancy migration barrier energies, obtained from the NEB calculations, are shown in Fig.~\ref{fig:GDist}. A Gaussian fit to this distribution yields a mean barrier of $0.81$ eV with a standard deviation of $0.32$ eV. The MPEA does indeed provide a rough energy landscape for vacancy migration, but to connect this to transport, we need more information than this, namely the extent to which this distribution is described by a random barrier (RB) or a random trap (RT) model. Figure~\ref{fig:NEB}a gives the result of one NEB calculation. 
The leftmost point is the energy of the initial state, the rightmost the energy of
the final state. 
The intermediate points are the energy of mid-transition
states, relaxed as the vacancy migrates, and the highest point corresponds to the saddle point of the transition. 

There are two primary effects of randomness in the distribution of barrier heights and well-depths. First, it affects the average transition rate. Second, it introduces correlations between subsequent hops. For example, in a purely RB model, a vacancy that just crossed a barrier with a small saddle energy $s$ is more likely to jump backwards over that same small barrier in the next hop. 
While the RT and RB models are constructed in terms of $s$ and $w$, they cannot be measured directly in a useful way. In the MPEA, there is no well-defined reference energy; both the $w$ before and $w$ after a transition are equally valid reference states, and many changes in the system (such as a swap between two atoms far away from the vacancy) can shift both $s$ and $w$ without meaningfully altering the transition barrier. Instead, we consider the symmetric and anti-symmetric components of the transition, $E_+ = (E_{A\rightarrow B}+E_{B\rightarrow A})/2$ and $E_- = (E_{A\rightarrow B}-E_{B\rightarrow A})/2$, respectively. If we choose a reference energy equal to the mean energy between the initial and final states, then
$s = E_+$ while $w_A = -w_B = E_-$. These are local descriptors of the transition barrier. If $s$ and $w$ are independent random variables, then the distributions of $E_-$ and $E_+$ are related to $w$ and $s$ in a predictable way (see Table~\ref{tab:stats}).

\renewcommand*{\arraystretch}{1.5}
\begin{table}[]
    \centering
    \begin{tabular}{|c|c|}
        \hline
        Quantity & $\text{var} (X)$ \\ 
        \hline
        $E = s -w$ & $\text{var} (s)+\text{var} (w)$\\ 
        \hline
        $E_+ = s - \frac{1}{2}(w_A + w_B)$ & $\text{var} (s) + \frac{1}{2} \text{var} (w)$ \\  
        \hline
        $E_- = \frac{1}{2}(w_B-w_A)$ & $\frac{1}{2} \text{var} (w)$ \\  
       \hline
    \end{tabular}
    \caption{Barrier properties and associated variance. These expressions are valid if $s$ and $w$ are independent, randomly distributed variables. \label{tab:stats}}
\end{table}

These relationships translate from global quantities like $s$ and $w$ to local quantities that directly compare the transitions between adjacent states. 
Distributions of $E_+$, and $E_-$ from the NEB calculations are given in Fig.~\ref{fig:NEB}b. A few initial observations can be made from this data. First, 
$\sigma_+ = 0.314 \pm 0.007$ eV, is significantly greater than 
$\sigma_- = 0.078 \pm 0.001$ eV. 
This is more consistent with an RB-like landscape, but there is still an RT component that cannot be ignored. There are also a substantial number of transitions with very small barriers ($E_+ \approx 0$). While states with ``zero barriers'' can be 
disregarded as unstable states, there are
a large number of states that are stable, but with very small non-zero barriers. This implies a small
barrier for the vacancy to hop between two sites in either direction. Intuitively, one can imagine a vacancy rapidly hopping back and forth between the two sites. 
As a result, \textbf{we should expect highly correlated vacancy migration and a theory for diffusion in this system must account for such correlations}.

These NEB calculations illustrate the nature of the energy landscape in this particular MPEA. However, it is not clear how $\sigma_w$ and $\sigma_s$ affect diffusion kinetics.  In the next section, we present Kinetic Monte Carlo results that predict the diffusion constant as a function of $\sigma_w$ and $\sigma_s$ and inform the development of a theoretical model.

\section{KMC Simulations}
\label{sec:kmc}
The NEB calculations provide the full distribution of transition energies $s$ and well depths $w$ in the CoNiFeCrMn MPEA. To understand the role of the distribution widths ($\sigma_s$ amd $\sigma_w$) on the vacancy diffusivity, we performed lattice Kinetic Monte Carlo (KMC) simulations. In KMC, we can directly specify the energy landscape and control the distribution of barriers. While $s$ and $w$ are not globally independent in real systems, we can construct a model system in which they are and use the relations in Table~\ref{tab:stats} to pick $\sigma_s$ and $\sigma_w$ such that $E_+$ and $E_-$ follow distributions observed in NEB calculations and reflect the local character of the MPEA. 

While setting up a traditional KMC simulation with a fixed distribution of $s$ and $w$ is trivial, the KMC set-up for MPEA vacancy diffusivity needs subtle, but important, modifications. These modifications are as follows:
\begin{enumerate}
    \item The energy landscape has to be dynamic. That is, one cannot \emph{a priori} assign the values of $s$ and $w$ on the lattice and keep them fixed during the KMC simulation. This can be easily seen by imagining a vacancy moving in a random solid-solution. If a vacancy starts at a site $i$, completes a few hops, and returns to its original site $i$, such that the local atomic configuration is modified by the swapping of neighbors, then the site and barrier energies also change. Therefore, a model that accounts for this dynamic energy landscape has to be considered.
    \item If the local atomic configuration does not change, then the site energy remains unchanged. That is, if a local atomic environment (consider the first nearest-neighbor shell around the vacancy) remains unchanged, 
    then the site energy should remain the same. 
    The same is true for the transition-state energies, except that we need to consider the environments of the both the initial and the final states.
\end{enumerate}

In this study, we developed an indexing approach (determined by the nearest-neighbor configuration) to guarantee that the dynamic nature and the translational symmetry of the energy landscape are satisfied. Please refer to Supplementary Information \ref{appendix:KMC} for a discussion the details of the KMC simulation set-up. It is important to mention here that our indexing approach does not satisfy invariance with respect to rotation/inversion symmetries. However, we do not anticipate this to change the results significantly because we are using a statistical model. By not considering the rotation/inversion symmetries, we simply add to the diversity of the configurations. As we keep the underlying distributions the same, adding the rotation/inversion symmetries would be computationally very expensive while not altering the results significantly.

Once the transition rates are calculated according to the $w$ and $s$ arrays, the
simulation proceeds via the Bortz-Kalos-Lebowitz algorithm \cite{bortz1975new}. 
For each transition $j$ from a site $i$, the rate is 
\begin{equation}
    r_{ij} = \nu \exp(-(s_{ij}-w_i)/{k_B T}),
\end{equation}
where $\nu$ is an attempt frequency,  $E_{ij} = s_{ij} - w_i$ is the migration barrier, and a transition is chosen randomly with a probability proportional to its rate. The vacancy site is then
occupied by the target neighbor atomic species and the neighbor is then occupied
by the vacancy and the time is incremented by
\begin{equation}
    \Delta t = \frac{1}{R} \ln(1/\epsilon),
\end{equation}
where $R$ is the sum of all transition rates and $\epsilon$ is a number chosen from a
uniform distribution $\epsilon \in (0, 1]$. This process is repeated for each step of
the simulation. The diffusion constant can then be computed by measuring the
mean-square displacement as a function of time, averaged over many runs of the
simulation (i.e., $\langle x^2 \rangle = 6 D t$). 

\begin{figure}[h!]
    \centering
    \includegraphics[width=0.6\linewidth]{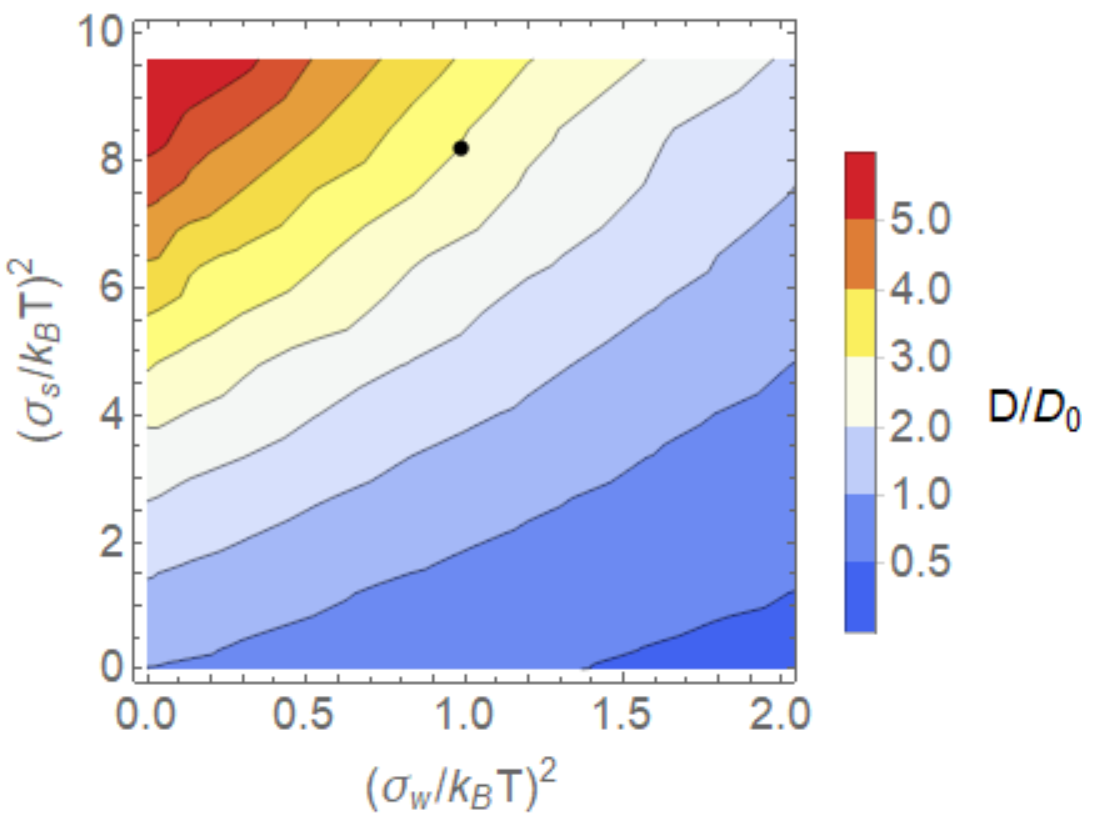}
    \caption{Diffusion constant $D$, computed via KMC simulation over a range of $\sigma_w$ and $\sigma_s$ at a temperature of 1273 K and scaled by $D_0$, the diffusion constant for a uniform barrier distribution equal to the mean barrier $0.81$ eV. The black circle denotes $\sigma_w$ and $\sigma_s$ chosen to match $\sigma_-$ and $\sigma_+$ as measured via NEB for the CoNiFeCrMn MPEA (see Table~\ref{tab:stats}).
    \label{fig:KMC}}
\end{figure}

The results of KMC simulations
at a fixed temperature $1273 K$ and mean barrier $\mu = 0.81$ eV (corresponding
to the NEB measurements) over a range of $\sigma_w$ and $\sigma_s$ are given in
Fig.~\ref{fig:KMC}. It is immediately apparent that while diffusivity decreases with increasing $\sigma_w$, it increases with increasing $\sigma_s$. For $\sigma_s$ and $\sigma_w$ chosen to match $E_+$ and $E_-$ for the CoNiFeCrMn MPEA, one would expect approximately a three-fold \textbf{enhancement} in vacancy diffusivity relative to a material with a uniform barrier distribution.
This makes intuitive sense; for a random trap model (large $\sigma_w$), the time is dominated by the vacancy's residence in deep energy wells. For a random barrier model, large barriers do not impede migration of the vacancy if there are alternative short barriers that it can traverse instead. 

This can be understood by analogy to resistors. For a random trap model, the roughness in the potential corresponds to a variation between subsequent hops; the total resistance of a series of resistors is given by the sum of resistances, which is dominated by the largest resistances. For a random barrier model, the roughness in the potential corresponds to a variation between alternate transitions out of a given state; the resistance of a set of resistors in parallel is given by the inverse of the sum of conductivities, for which small resistances dominate. One cannot decrease resistance by adding a weak resistor in series no more than one can increase the resistance by adding a strong resistor in parallel. To fully explain these results and make future predictions, we next present a statistical theory of vacancy migration in disordered energy landscapes.
\section{Theory} \label{sec:theory}

The KMC simulations provide the foundational observation that random trap landscapes impede diffusion while random barrier landscapes enhance it. An analytical theory is necessary to explain these observations and make intuitive predictions. A rough energy landscape can have two effects on diffusion. First, it can change the average transition rate $\Gamma$ (analogous to the effective rates computed in prior work \cite{zwanzig1988diffusion,mussawisade1997combination,banerjee2014diffusion,seki2015relationship,seki2016anomalous}). Second, it can introduce correlations between hops. Consider a rough landscape given by random, independent distributions of well-energies $w$ and transition saddle-point-energies $s$, with means $\mu_w$ and $\mu_s$ and standard deviations $\sigma_w$ and $\sigma_s$. The diffusion constant $D$ can be expressed as
\begin{equation}
    \frac{D}{D_{\circ}}= \frac{\Gamma(\sigma_s,\sigma_w,\mu) \mathbb{F}(\sigma_s,\mu)}{\Gamma_{\circ}},
    \label{eq:Drat}
\end{equation}
where $D_\circ$ and $\Gamma_\circ$ are the diffusion constant and jump frequency for a  random walker in a landscape with constant vacancy migration barrier energy $\mu = \mu_s-\mu_w$ and $\mathbb{F}$ is the effect of correlations. When the vacancy migration barrier is a constant value, there are no correlations and $\mathbb{F} = 1$. It is important to note that $\mathbb{F}$ depends only on the distribution of saddle-point energies and not on the distribution of well-energies. That is, for a purely random trap energy landscape $\mathbb{F} = 1$. This is because $w$ is the property of the lattice site and all jumps out of that site are equally likely (refer to Figure \ref{fig:rtb}(a)). Therefore, the randomness in the well-energies will not affect which transition is selected and hence cannot influence correlations between subsequent hops. Therefore, we omit the dependence of $\sigma_w$ for $\mathbb{F}$ in equation \ref{eq:Drat}. In the following, we will derive the analytical form for the jump frequency $\Gamma(\sigma_s,\sigma_w,\mu)$ and the effect of correlations $\mathbb{F}(\sigma_s,\mu)$ in the disordered energy landscape. The derivation will include the following steps:

\begin{itemize}
    \item First, we will show that the contributions of the disorder in well-energies can be separated from those in the saddle-point energies. That is, the mixed model can be simplified into two independent contributions arising from a pure random-trap and a pure random-barrier model.
    
    \item The random trap contribution is equivalent to the expression derived by Zwanzig \cite{zwanzig1988diffusion}. For the random-barrier contribution, we outline the steps required to compute the average transition rate $\Gamma(\sigma_s, \mu)$ for a random barrier model. The key insight that we will develop here is that the distribution of barriers-crossed (hops) is not the same as the distribution of barriers present in the system.
    
    \item Finally, we derive the effect of correlations $\mathbb{F}(\sigma_s,\mu)$ in the random barrier model. As the well-energy distributions do not contribute to $\mathbb{F}$ (as discussed above), it suffices to simply compute the correlations for the pure RB model.
\end{itemize}

\subsection{Separation of the random trap (RT) and random barrier (RB) contributions}

Assuming vacancy migration is thermally-activated with an attempt frequency $\nu$, the rate of a transition, from state $i$ to $j$, is given by $\Gamma_{ij} = \nu \exp(-E_{ij}/k_B T)$, where $E_{ij} = s_{ij} - w_i$. The average time $\langle \tau \rangle$ that a vacancy resides in state $i$ is therefore

\begin{equation}
    \langle \tau \rangle = \left \langle \left(\sum_j^z  \nu e^{-\frac{E_{ij}}{k_B T}}\right)^{-1} \right \rangle
    = \nu^{-1} \left \langle  e^{\frac{-
    w_i}{k_B T}} \left(\sum_j^z e^{-\frac{s_{ij}}{k_B T}}\right)^{-1} \right \rangle,
\end{equation}


where $z$ is the coordination number and the number of possible jumps for the vacancy. If $w_i$ and $s_{ij}$ are independent random variables, their corresponding terms can be separated. That is, 


\begin{align}
\Gamma= \langle \tau \rangle^{-1} = \frac{\nu \langle \exp(-w_i / k_B T) \rangle^{-1}}{\left \langle \left( \sum_j^z \exp(-s_{ij}/k_B T) \right)^{-1} \right \rangle}.
\end{align}

The average $\langle \rangle$ is over all \textbf{hops} (i.e,. all chosen transitions), rather than all possible transitions. As the choice of a transition is independent of $w_i$, an average over $w_i$ encountered by the vacancy is simply the system-wide normal distribution of $w$ and the Arrhenius part of the RT component is

\begin{equation}
    \mathbb{A}_{RT} = \left\langle \exp{\left(- \frac{w_i}{k_B T} \right)} \right\rangle^{-1} = \exp \left(\frac{\mu_w}{k_B T} \right) \exp \left[- \left( \frac{\sigma_w}{k_B T} \right)^2 \right], \label{eq:GammaRT}
\end{equation}

identical to Zwanzig's result \cite{zwanzig1988diffusion}. Here, we introduce the symbol $\mathbb{A}$ for the Arrhenius part of the jump frequency. The remaining RB component,
\begin{equation}
    \frac{\Gamma}{\mathbb{A}_{RT}} = \nu \left\langle  \left[ \sum_{j=1}^z \exp \left( -\frac{s_{ij}}{k_B T} \right) \right]^{-1}  \right \rangle ^{-1} = \nu \mathbb{A}_{RB},
    \label{eq:GammaRB}
\end{equation}
is independent of $w$. That is, we can solve the case of the mixed energy model as independent contributions of a pure random-trap model, $\mathbb{A}_{RT}$ (Eq. \ref{eq:GammaRT}), and a pure random-barrier model $\mathbb{A}_{RB}$ (Eq. \ref{eq:GammaRB}), that is $\Gamma = \nu \mathbb{A}_{RT} \mathbb{A}_{RB}$. 

\subsection{Computing the jump frequency for a pure random-barrier model}

To compute the arrhenius part, $\mathbb{A}_{RB}$ in equation \ref{eq:GammaRB}, we have to determine the distribution of the barriers crossed by the vacancy in the random-walk trajectory. This is different from the distribution of the barriers present in the system as the vacancy prefers to jump over smaller barriers. For a uniform landscape, where the migration energy barrier is given by $\mu = \mu_s - \mu_w$ and $\sigma_w =\sigma_s = 0$, the Arrhenius parts of the jump frequency are given by $\mathbb{A}_{RB}^{\circ} = \exp(-\mu_s/k_B T)$ and $\mathbb{A}_{RT}^{\circ} = \exp(\mu_w/k_B T)$. The jump frequency, in the uniform barrier case, has the form $\Gamma_{\circ} = \nu \mathbb{A}_{RT}^{\circ} \mathbb{A}_{RB}^{\circ} = \nu \exp(-\mu/k_B T)$. That is, while $\mathbb{A}_{RB}$ and $\mathbb{A}_{RT}$ are sensitive to the choice of reference energies (e.g., choosing $\left( \mu_s, \mu_w \right) = \left( \mu, 0 \right)$ or $\left( \mu_s, \mu_w \right) = \left( 0, -\mu \right)$),  the product $\mathbb{A}_{RB} \mathbb{A}_{RT}$ is not. In the disordered energy landscape, $\mathbb{A}_{RB}$ is independent of the distribution of well-energies, $w$, and it can be calculated by fixing $w$ to a constant value, i.e. $w = \mu_w$. For clarity, we choose a reference energy such that $\mu_w=-\mu$ and $\mu_s = 0$ and adopt a reduced notation $u = s/k_B T$, $u_0 = -\mu/k_B T$, and $x = \left(\sigma_s/k_B T \right)^2$. $u$ is drawn from a truncated normal distribution:
\begin{equation}
P_1(u,\mu,x) = N_1(\mu,x) e^{-\frac{u^2}{2 x}}, \label{eq:p0}
\end{equation}
over the domain $u \geq u_0$. $N_1(\mu,x)$ is a normalization constant. From a given lattice site, the probability $p$ of selecting a given transition is 
\begin{equation}
p = \frac{e^{-u}}{e^{-u} + \sum^{z-1}_j e^{-v_j}}, \label{eq:pi}
\end{equation}
where $u$ is the barrier of the target transition and $v_j$ represent all of the other transition barriers.

\begin{figure}[h!]
\centering
    \includegraphics[width=.6\linewidth]{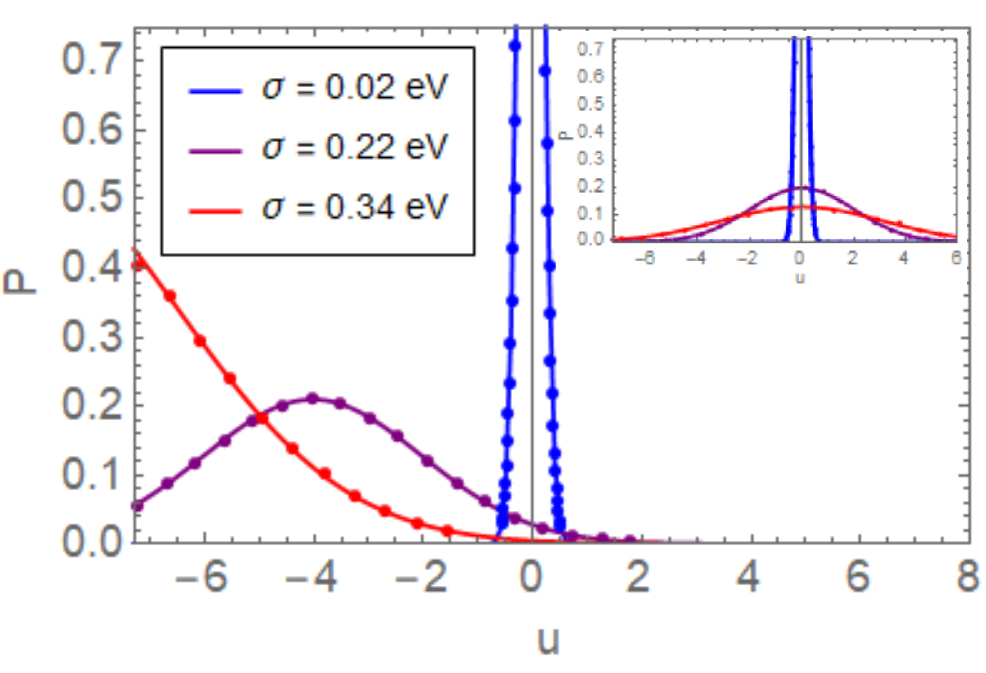} \\
\caption{$P_c$ and $P_1$ (inset) distributions, as computed by theory (solid) and from KMC (points). The distributions are plotted as a function of the barrier energy $u = s/k_B T$.
}
\label{fig:prob1}
\end{figure}

If the vacancy jumps over a small barrier $u$, it is likely to hop backwards across that same barrier (for the \textbf{random-barrier} energy landscape). This means that the distribution of $u$ encountered by the vacancy should be biased towards smaller barriers relative to the system-wide barrier distribution, $P_1$. The distribution of \textbf{chosen} barriers $P_c(u)$ is
\begin{equation}
P_c(u) = P_1\sum^\infty_{i=1}p^i = N(x,u_0) e^{-\frac{(u+x)^2}{2x}}, \label{eq:pc}
\end{equation}
where $N(x,u_0)$ is a normalization constant. Equation \ref{eq:pc} can be understood as follows: the probability that a barrier first encountered by the vacancy has height $u$ is given by $P_1$ (equation \ref{eq:p0}). The probability of the vacancy crossing that barrier once is given by $p$ (equation \ref{eq:pi}), crossing twice is given by $p^2$, etc. The sum of these probabilities is then $P_c$. This distribution (with $P_1$ inset), compared with KMC results, is shown in Fig.~\ref{fig:prob1}. The $P_c$ distribution is nearly identical to the KMC results. 

The Arrhenius part of the random barrier model, $\mathbb{A}_{RB}$, is related to the inverse of the average of residence times (as shown in equation \ref{eq:GammaRB}). That is, the jump frequency for the RB model can be expressed as $\Gamma_{RB} = \nu \mathbb{A}_{RB}$ where $\mathbb{A}_{RB} = \left[ \nu \langle \tau \rangle \right]^{-1}$. Here the residence time, $\tau$ of the vacancy in the lattice-site $i$ is given by:

\begin{equation}
\nu \tau  = \left[ \sum_{j=1}^z \exp \left( -\frac{s_{ij}}{k_B T} \right) \right]^{-1}
\end{equation}

\noindent where $s_{ij}$ is the saddle-point energy for the hop from state $i$ to state $j$. Suppose the vacancy arrives at state $i$ by hopping over a barrier given by the non-dimensionalized variable $u$. The barriers that the vacancy now observes, from state $i$, are $u$ and $v_{j}$ ($j=1,\ldots, z-1$). We use two different variables $u$ and $v$ to denote the barriers from state $i$ to emphasize the fact that they are drawn from two distinct probability distributions. Since the vacancy just hopped over the barrier $u$, the corresponding distribution is given by $P_c$ (equation \ref{eq:pc}. The distributions of the remaining $z-1$ barriers, defined by random variables $v_{j}$, correspond to the system-wide function, $P_1$ (equation \ref{eq:p0}). The residence time, $\tau$, can be re-written in-terms of the the non-dimensional variables, $u$ and $v_{j}$, as:

\begin{equation}
\nu \tau =  \left[ \exp \left( -u \right) + \sum_{j=1}^{z-1} \exp \left( -v_{j} \right) \right]^{-1}    
\end{equation}

The mean of the residence times, $\nu \langle \tau \rangle$, can be obtained by a twelve-fold integral of the rate terms, $e^{-u}$ and $e^{-v_{j}}$, multiplied by the appropriate distribution functions, $P_c$ and $P_1$, respectively. Unfortunately, there are no analytical solutions for the integration of the rate terms, $e^{-v_{j}}$, when the random variable $v_{j}$ is described by a truncated normal distribution.

Instead, we introduce approximate probability distributions for the summation of rates, defined using the variable $r_k = \sum_{j=1}^k \exp \left( - v_j \right)$, where $v_j$ is a random variable drawn from a truncated normal distribution. $r_k$ is now a random variable, defined as the sum of rates over $k$ barriers, with a probability distribution $\rho_k \left( r_k \right)$. The determination of the distributions, $\rho_k$, is described in  Supplementary Information~\ref{appendix:lognormal}. Equipped with the distribution functions, $\rho_k$, we can now simplify the computation of the average residence time $\nu \langle \tau \rangle$. The residence time, for site $i$, can be re-written as $\nu \tau =  \left(e^{-u} + r_{z-1} \right)^{-1}$ and the average residence time can simply be computed as a two-fold integral over the random variables $u$ and $r_{z-1}$, multiplied by the appropriate distribution functions $P_c$ and $\rho_{z-1}$, respectively. Therefore, the Arrhenius part, $\mathbb{A}_{RB}$, can be written in terms of a two-fold integral as:
\begin{equation}
    \mathbb{A}_{RB} = \left[ \nu \langle \tau \rangle \right]^{-1} = \left[ \iint \left(e^{-u} + r_{z-1} \right)^{-1} P_c(u,\mu,\sigma) \rho_{z-1}\left( r_{z-1} \right) du dr \right]^{-1},
\end{equation}
\noindent which can be evaluated numerically. As mentioned above, the barrier that the vacancy most recently hopped is drawn from the $P_c$ distribution (Figure \ref{fig:prob1}). As $P_c$ favors smaller barriers, the average rate in the $RB$ model will be higher than that for a uniform distribution. This also implies correlations between subsequent hops - if the vacancy hops over a particularly short barrier, it is more likely to hop backwards in the direction opposite the previous hop. 

\subsection{Computing the correlation factor between hops}

The $j$th correlation factor $f_j = \langle \vec{v}_i \cdot \vec{v}_{i+j} \rangle$ is the average cosine between a given hop $i$ and $j$ hops subsequent. If the hops are completely uncorrelated, then $f_j=0$. In a vein similar to the derivation of the $\mathbb{A}_{RB}$, we will suppose that a vacancy reaches site $i$ by hopping over a barrier $u$. Let us also denote the barrier over a hop in the same direction as the previous jump as $v$. Then the average cosine for the next hop from site $i$ is given by:

\begin{equation}
\alpha_{1} (u,v, r_{z-2}) = \frac{e^{-v}-e^{-u}}{e^{-v}+e^{-u} + r_{z-2}}, \label{eq:alpha1}
\end{equation}

\noindent where $r_{z-2}$ is the sum of the rate terms of the remaining barriers, i.e. $r_{z-2} = \sum_{j=1}^{z-2} e^{-v_{j}}$. This expression can be rationalized as follows: (i) the $e^{-u}$ term reflects the possibility that the vacancy hops backwards over the same barrier $u$ (corresponding to a cosine of $-1$), (ii) the $e^{-v}$ term corresponds to a hop in the same direction as the previous hop (corresponding to a cosine of $1$), and on average, the cosine terms of all other hops cancel out in the numerator. (iii) In the denominator, we have the summation of the probabilities of all possible jumps, given by $e^{-u}$, $e^{-v}$ and $r_{z-2}$ (for the remaining $z-2$ jumps). As discussed in Supplementary Information~\ref{appendix:lognormal}, $r_{z-2}$ is a random-variable whose distribution is given by the function $\rho_{z-2}$.

\begin{figure*}[h!]
\centering
\begin{tabular}[b]{c}
    \includegraphics[width=.44\linewidth]{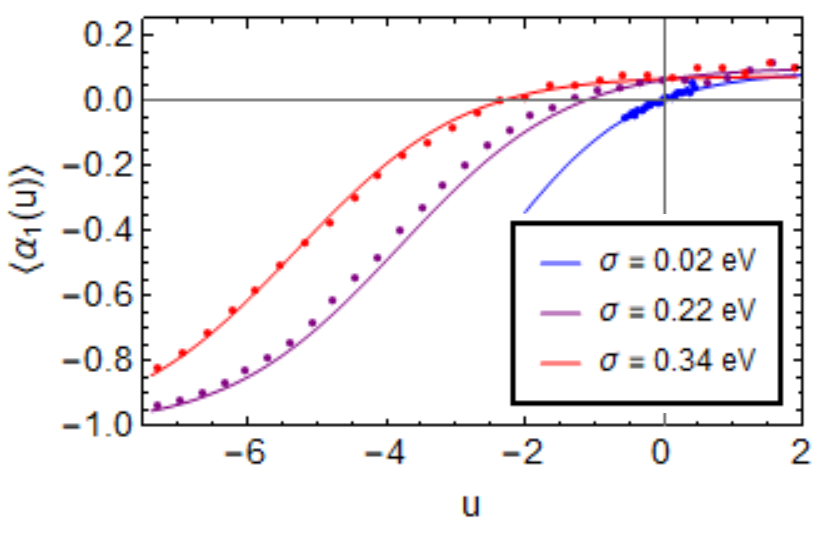} \\
    \small (a) $\langle \alpha_1 (u) \rangle$
  \end{tabular} \quad
\begin{tabular}[b]{c}
    \includegraphics[width=.46\linewidth]{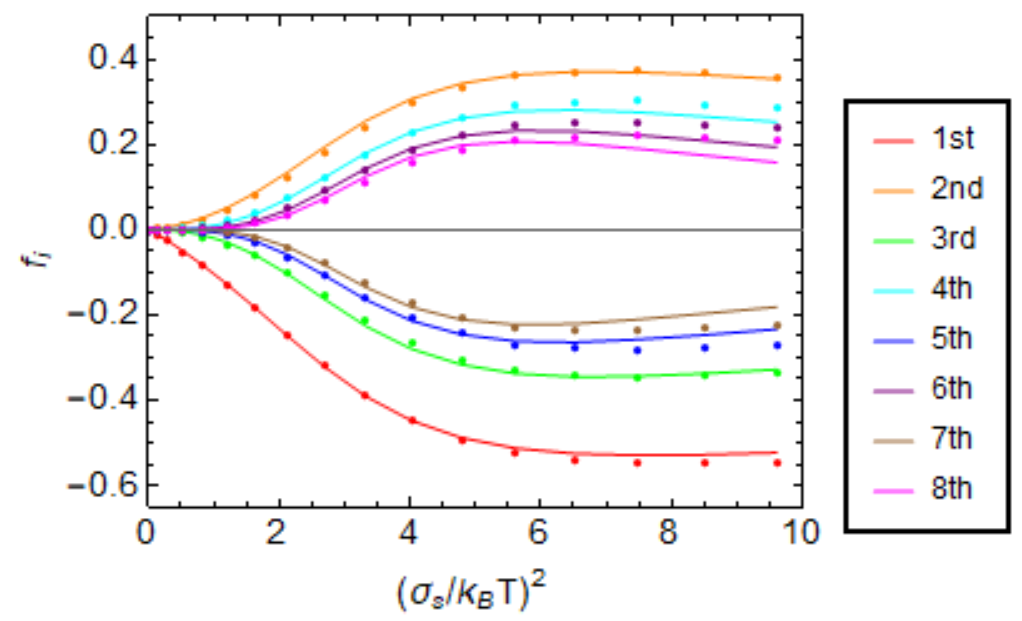} \\
    \small (b) Correlation Factors
  \end{tabular} 
\quad
\caption{a) Marginal distribution $\langle \alpha_1(u) \rangle$ as a function of previously-hopped barrier $u$ for different values of $\sigma_s$, compared with measurements from KMC (points). b) The correlation factors $f_i$ in the random-barrier model. \label{fig:corrAlpha}}
\end{figure*}

The first correlation factor $f_1$ is then obtained by integrating over the distributions $u$, $v$ and $r_{z-2}$ as:
\begin{align}
f_1 = \iiint \mathbb{P}(u,v,r_{z-2},x) \alpha_1 (u,v,r_{z-2}) du dv dr \label{eq:f1}, \end{align}
where 
\begin{align}
\mathbb{P}(u,v,r_{z-2},x) = P_c(u,x) P_1(v,x) \rho_{z-2}(r_{z-2},x). \label{eq:PP}
\end{align}
These integrals can be easily evaluated using numerical integration techniques. In Figure \ref{fig:corrAlpha}(a), we plot the marginal distribution $\langle \alpha_1 (u) \rangle$ defined as:
\begin{equation}
\langle \alpha_{1} (u) \rangle = \iint P_1(v,x) \rho_{z-2}(r_{z-2},x) \alpha_1 (u,v,r_{z-2}) dv dr, \label{eq:alpha1_mar}
\end{equation}
Shown in Figure \ref{fig:corrAlpha}(a), is a close match between $\langle \alpha_{1} (u) \rangle$ and the KMC simulations for three different values of $\sigma_s$. Note the convergence of $\langle \alpha_1 (u) \rangle$ to $-1$ as $u$ decreases. For smaller barriers $u$, it is very highly likely that the vacancy will jump back over that small barrier. This results in an angle of 180$^{\circ}$ between subsequent hops and the average cosine will converge to -1.


For higher order correlations $f_i$ (i.e., the correlation between a hop and the $i$th subsequent hop), one must determine corresponding $\alpha_i$ terms (see Supplementary Information 3). The first eight correlation factors ($f_1, f_2, \ldots, f_8$) are plotted in Fig.~\ref{fig:corrAlpha}b, compared directly with measurements from KMC simulations. The theory is nearly an exact match to the simulations, except for the high-order correlation factors at large $x$. This is discussed further in Supplementary Information \ref{sec:fcorr}. The effect of correlation factors on the diffusion coefficient, up to the n$th$ correlation factor, is given by
\begin{equation}
    \mathbb{F} = 1 + 2 \langle v_i \cdot v_{i+j} \rangle = 1 + \frac{2}{n}\sum_j^n (n-j+1) f_j.
\end{equation}
This is a modification of the common random walker solution with correlated hops for finite sums \cite{shewmon2016diffusion}; if, for example, we consider up to the 8th correlation factor, then we sample from an initial hop up to 9 hops in the future. This means that 8 instances of $f_1$ count toward the average, 7 instances of $f_2$, etc. We note here that it is common in the theory of correlated random walks to approximate $\mathbb{F}=(1+f_1)/(1-f_1)$ \cite{shewmon2016diffusion}; this is only valid if every state is equivalent, such that $f_j = f^j_1$. That is not true in this case, as evidenced by Fig.~\ref{fig:corrAlpha}.

\section{Results and Discussion}

\begin{figure*}[h!]
\centering
\begin{tabular}[b]{c}
    \includegraphics[width=.45\linewidth]{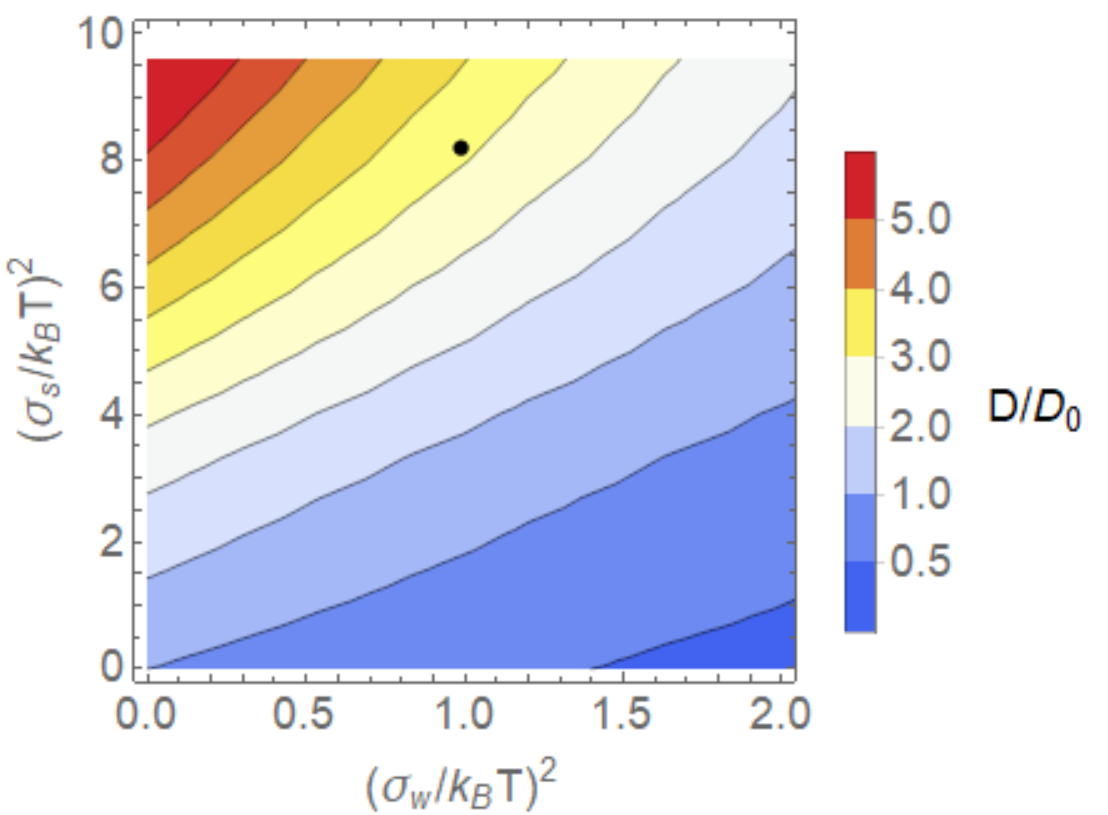} \\
    \small (a) Random-Walk Theory
  \end{tabular} \quad
\begin{tabular}[b]{c}
    \includegraphics[width=.45\linewidth]{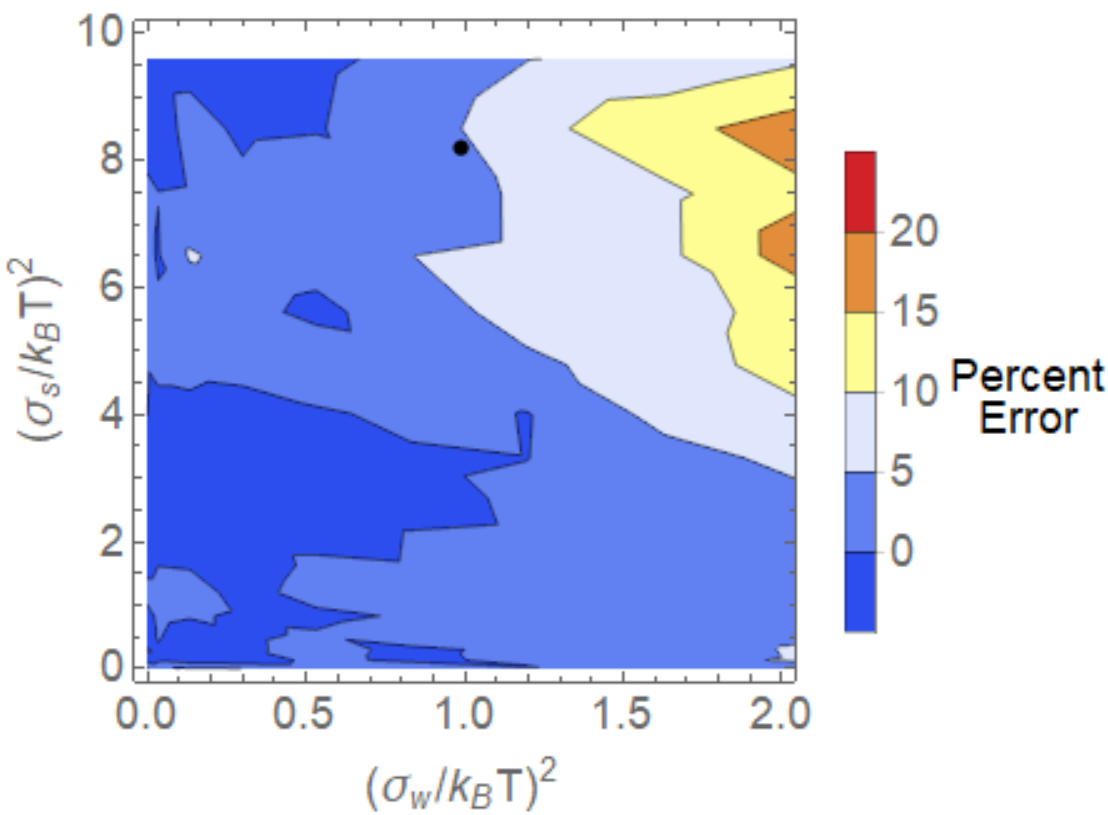} \\
    \small (b) Error
  \end{tabular}
\quad
\caption{a) $D/D_\circ$ as calculated via analytical theory with mean barrier $\mu = \mu_s-\mu_w = 0.81$ eV and temperature 1273 K. b) Percent error between analytical theory and KMC. $D$ is the diffusion constant, while $D_\circ$ is the constant for a landscape with uniform barriers of height $\mu$. The black circle denotes $\sigma_w = 0.11$ eV and $\sigma_s = 0.30$ eV, corresponding to the $E_+$ and $E_-$ distributions measured for the CoNiFeCrMn MPEA, as measured by NEB.
\label{fig:theoryError}}
\end{figure*}

The analytical diffusivity (the ratio $D/D_{\circ}$, where $D_{\circ}$ is the diffusivity when $\sigma_w = \sigma_s = 0$) is plotted as a function of $\left( \sigma_w/k_B T
\right)^2$ and $\left( \sigma_s/k_B T\right)^2$ in Fig.~\ref{fig:theoryError}a, and the error between the theory and KMC is given in Fig.~\ref{fig:theoryError}b. The theory and KMC produce nearly identical results, except when both $\sigma_w$ and $\sigma_s$ are large. This is because while the random barrier model does not allow the $s$ component of transition energy to be less than $\mu_w$, it does not account for the effect of particularly deep wells, which could allow stable configurations where $s<\mu_w$ as long as $s>w$. This is, however, only a problem when $\sigma_w$ and/or $\sigma_s$ are much greater than measured for the CoNiFeCrMn HEA. 

These plots show that vacancy diffusivity can either be retarded
($D/D_{\circ} \sim 0.5$ if $\sigma_w = k_BT\sqrt{2}$ and $\sigma_s = k_B T$)  
or accelerated
($D/D_{\circ} \sim 2$ if $\sigma_w = k_B T$ and $\sigma_s = \sqrt{5} k_B T$) depending on the
widths of the distributions of site energies $(w)$ and transition-state energies $(s)$. This captures the full breadth of behaviors observed in more recent experimental results, while perhaps explaining the assumed universality of sluggish diffusion in MPEAs. The KMC simulations and theoretical model presented here provide insight to the relationship between the energy landscape and the resulting transport properties. In the most general terms, a large $\sigma_s$ (or $\sigma_+$) yields enhanced diffusion, while a large $\sigma_w$ (or $\sigma_-$) produces sluggish diffusion. 
Diffusion constants may be difficult to measure directly, but the method presented here, to compute  $E_+$ and $E_-$ distributions, is amenable to automated survey that can be used to probe the high-dimensional compositional space of MPEAs. 

\begin{figure}[h!]
    \centering
    \includegraphics[width=0.6\linewidth]{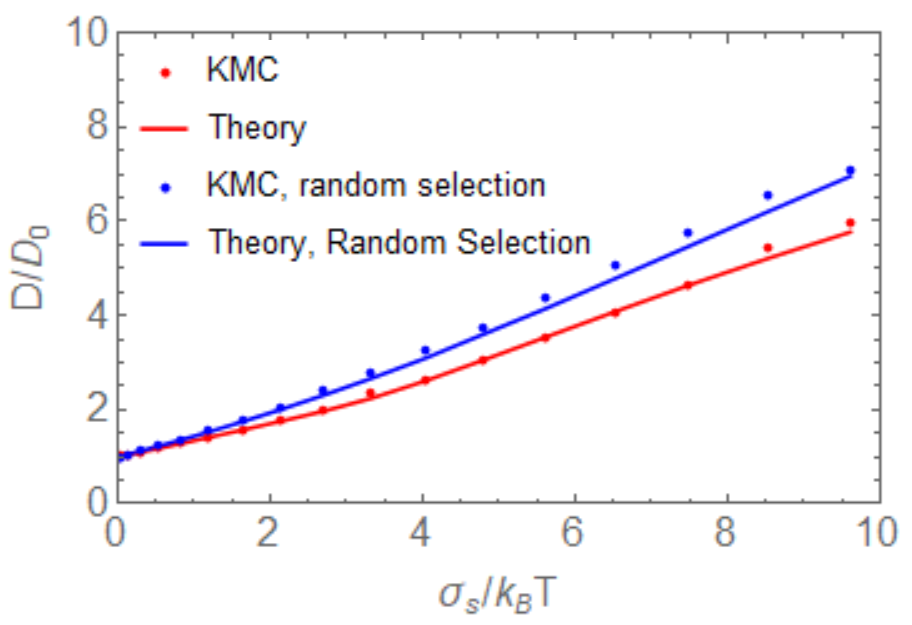}
    \caption{Diffusion constant for theory and KMC (red) for a pure RB model, compared with a model in which all transitions have equal probability (blue). For these calculations, $\mu = 0.81$ eV.}
    \label{fig:uncorr}
\end{figure}

The model also allows us to directly quantify the influence of correlations between hops and barrier selection on diffusion in a random energy landscape. Figure~\ref{fig:uncorr} shows a comparison between KMC and Theory for the pure random barrier model and a simplified model. In the simplified KMC, all transitions from a given state are given equal probability $1/z$, irrespective of the barrier energy, at each time step. In the simplified theory, we equate the probability of hopped barriers to the system-wide distribution of transition state energies, i.e. $P_c=P_1$. Under these conditions, the averaging over all chosen vacancy hops (Eq. \ref{eq:GammaRB}) can be computed by simply averaging over the entire energy landscape, making it more similar to the Effective Medium approximation common in existing literature \cite{webman1981effective,ansari1985protein,haus1987diffusion,ambaye1995asymptotic,mussawisade1997combination,seki2016anomalous}. The two cases are very similar due to the competing effects of preferential small-barrier selection and hop correlation. For example, when $\sigma_s = 0.34$ eV,  $\Gamma$ is approximately $\sim 2.3$ times larger than in the equal-probability case but the correlation factor $\mathbb{F} \approx 0.3$. Therefore, the difference between the two scenarios (as shown in Figure \ref{fig:uncorr}) is not very large. However, this is not guaranteed for different temperatures and mean barriers. It is important to identify the correct theory for modeling the random energy landscape if we wish to accurately model diffusion in disordered materials.  

Irrespective of the assumptions that went into this KMC and theoretical model, it is clear that the random barrier/trap character of the energy landscape can profoundly influence diffusion, potentially altering the effective diffusion constant by at least an order of magnitude; these details are necessary to predict diffusion behaviors in disordered systems and must be considered in future models.

\section{Conclusions} 
\label{sec:conc}

The set of experimentally synthesized alloys cover only a small region of the high-dimensional compositional space that has been identified as thermodynamically plausible for MPEAs \cite{senkov2015accelerated, miracle2017critical}. To sample this high-dimensional space in an efficient manner and to design novel MPEAs with targeted properties, the need for new theoretical and computational tools has been highlighted \cite{george2019high}. 
In this current study, we focused on developing analytical tools for predicting diffusion kinetics, which are time consuming to measure either using experiments or atomistic simulations. For diffusivity in MPEAs, the computational tools require a foundation of a) characterization of the kinetic barriers to diffusion, b) a means of simulating diffusion in a model system faithful to the observed energy landscape, and c) a theoretical framework for understanding how the underlying energy landscape relates to diffusion. 

In this article, we presented a complete cross-section of diffusion in rough energy landscapes, where the disorder is representative of solid-solution  MPEAs.
We developed a flexible KMC protocol where the distributions of the disordered well and the saddle-point energies can be independently controlled.
To better understand the KMC results and to develop predictive models, we presented a theoretical framework for vacancy diffusivity in disordered energy landscapes.
While the statistics used in the KMC simulations are informed by the direct computation of the migration barriers for a CoNiFeCrMn EAM potential, the developed theory spans a wide-range of distributions that one may observe in generic MPEAs. Therefore, for a given alloy system, equipped with knowledge of the well and saddle-point energy distributions, simulations and theoretical results provided in this article can be used to predict transport properties. The theory itself lends to an intuition regarding transport; wider distributions of saddle-point energies enhance diffusion while wider distributions of well depths stifle diffusion. If these distributions can be connected to alloy chemistry and compositions, designer MPEAs with optimized transport properties can be achieved.


As it stands, our model is insensitive to correlations in the energies. That is, the relationship between neighboring species and migration barriers is completely random. There is also no mixing enthalpy and no consideration for different migration barriers for different atomic species. This is significant; it has been noted, for example, that the presence of Mn is associated with sluggish diffusion in the CoNiFeCrMn MPEA \cite{dkabrowa2019demystifying}. Our own observations suggest that Mn generally has the smallest barrier to migration, consistent with the findings of Tsai et al. \cite{tsai2013sluggish}. This seemingly contradictory result highlights the subtle nature of this effect and the need for tools to directly measure the energy landscape and connect that to actual diffusion behavior. While our model does not address the influence of different species having different kinetics (which our NEB data does suggest), existing Onsager-type continuum and multi-frequency models \cite{moleko1989self,druger1983dynamic,perondi1997bond,perondi1997trap,allnatt2016high,trinkle2018variational} may be used in conjunction with our method to address this.
In future work, the tools presented here will be augmented appropriately to investigate these effects and guide the design of multi-principal alloy systems.

We hope that this statistical treatment of vacancy diffusion provides a building block for future navigational tools in the vast MPEA sea. In the near term, it should also serve as a guide for what computational tools currently in development must do correctly. The potential that we used \cite{choi2018understanding} is a laudable effort - a necessary step for the atomistic modeling of these complex alloys. However, for understanding vacancy diffusivity, it is insufficient to match just the vacancy migration barriers. That is only half the information; holding that steady, one can change the diffusion constant by orders of magnitude. MPEA potentials cannot be relied upon for diffusion-related problems if they do not capture the full nature of the vacancy migration landscape - the symmetric and anti-symmetric barrier components used here may be useful quantities for measurements and fits. This necessary level of detail may also be an opportunity for novel machine learning approaches to modeling transport in MPEAs \cite{kostiuchenko2019impact}.

\section*{Acknowledgements}
This work is supported by the U.S. National Science Foundation under Grant No. CMMI-1826173.


\newpage
\clearpage

\section*{References}
\bibliography{main}
\bibliographystyle{elsarticle-num}

\newpage

\pagenumbering{arabic}
\setcounter{page}{1}
    


\setcounter{section}{0}
\renewcommand{\thesection}{S\arabic{section}}
\renewcommand{\theHsection}{S\the\value{section}}
\setcounter{figure}{0}
\renewcommand{\thefigure}{S\arabic{figure}}



\section{KMC Details} \label{appendix:KMC}

For the initial configuration, we generate an FCC lattice of sites, randomly populated with one of five atomic species. One site is populated with a vacancy. The rejection-free Kinetic Monte Carlo algorithm we use is standard \cite{bortz1975new}. However, the choice of saddle points and well-depths for transitions must be done carefully and efficiently. We chose our algorithm to satisfy the following constraints: 
\begin{enumerate}
\item \textbf{Translational symmetry}: the same set of nearest neighbors results in the same barriers, regardless of where it is in absolute space. 
\item \textbf{Well-sampled barriers}: many different barriers are represented and there are few identical barriers except via translational symmetry. 
\item \textbf{Zero mixing enthalpy}: no species are preferential neighbors. This term could be added subsequently to introduce short-range order, but this effect should be independently controllable.
\end{enumerate}
The requirement of translational symmetry is satisfied if the barriers are a function of the nearest neighbor environment. Good sampling, as well as the zero mixing enthalpy requirement imply that bond models, or simply counting the number of neighbors of each species of the vacancy, would be insufficient. Additionally, the saddle point $s$ should be:
\begin{enumerate}
    \item \textbf{Symmetric}: the transition saddle-point energy from configuration A to configuration B should be the same as the energy from B to A.
    \item \textbf{Uncorrelated}: the $z$ saddle point energies of transitions from a given state should be independent - such correlations could be introduced in future work, but this should be independently controllable. 
\end{enumerate}

To satisfy these conditions, we initially generate random arrays of saddle heights $s$ and well depths $w$ with corresponding distribution widths $\sigma_s$ and $\sigma_w$, as well as mean barrier heights $\mu_s$ and $\mu_w$ ($\mu_w$ is chosen as zero and $\mu_s$ is chosen as $\mu$, as the reference energy does not matter). If $s<\mu_w$ or $w>\mu_s$, they are rejected and a new energy is generated randomly from the normal distribution. This is to prevent the frequent drawing of unphysical negative transition barriers. As long as $\sigma_w$ and $\sigma_s$ are sufficiently smaller than the difference between $\mu_s$ and $\mu_w$, this has a negligible effect on the simulation.

At each KMC step, the species of sites neighboring the vacancy site A are used to generate an index $i$ for accessing the $w$ array (i.e., $w$ is the same for every transition out of a given state). For the $w$-index $i$, we consider only the $z$ nearest neighbors of the vacancy site:
\begin{equation}
    i = \sum_{l=1}^z t_l (m+1)^{l-1}, \label{eq:qIndex}
\end{equation}
where $m$ is the number of different species and $t_l$ is the type of a given neighbor (numbered from 0 to $m$; for the CoNiFeCrMn alloy, we choose 0 for cobalt, 1 for nickel, etc.). Here, the neighbors are sorted such that the neighbor $(0,1/2,1/2)$ from the central atom is always the $1^\text{st}$ neighbor, $(0,1/2,-1/2)$ always the $2^\text{nd}$, etc. Every possible configuration of neighbor species generates a different index.

A separate index, $j$, is determined to reference $s$ for each of the z transitions available to the vacancy (corresponding to the nearest-neighbor sites). $j$ is computed using the neighbors of \textbf{both} the vacancy site A and the destination site B: 
\begin{equation}
    j=  \sum_{l=1}^{2z-c} t_l (m+1)^{l-1} \mod \phi \label{eq:hIndex1},
\end{equation}
where $c$ is the number of common neighbors between any two pairs of neighbors (4 for FCC, making for 20 total terms, including the origin and destination site). We take the modulo $\phi$ because $j$ would otherwise occupy a very large range from 0 to approximately $ 6^{20}$, most of which would never be realized. In the interest of computational efficiency (and to prevent integer overflow), we take the modulo $\phi$, where $\phi = 7^7$. This choice of $\phi$ is large and coprime with $(m+1)$, minimizing the number of overlapping indices. 
The summation terms $t_l(m+1)^{l-1} \mod \phi$ can be pre-calculated and stored in a $20\times6$ array, removing the need for unnecessary computation.  To enforce the transition-reversal symmetry of the $s$ term, this index is computed for the forward and reverse transition and the \textbf{indices} are added together to produce the final index used to access the $s$ array.

It should be noted that this formulation lacks any rotation/mirror/inversion symmetries. Each neighbor is unique with respect to the lab frame; shifting all atomic species and the vacancy one site over will preserve the transition barriers, but rotating all species around the vacancy will not. This sacrifice saves substantial computing time as it allows each neighbor configuration to map to an array index. Due to the disordered random nature of the transition barriers and the size of the neighbor configuration space, we do not expect this to influence the kinetics in any meaningful way. However, the effect of rotational/inversion symmetry could be a subject of further inquiry.
\section{Determination of distributions of the sum of transition rates} 
\label{appendix:lognormal}
In the analytical theory, we are often tasked with computing the sum of multiple  independent, randomly-distributed rates (e.g., Eqs.~\ref{eq:pi}, \ref{eq:alpha1}, \ref{eq:f1}, and \ref{eq:PP}). In theory, one could integrate over these variables independently, but this would entail a 12-fold integral that is difficult and expensive to evaluate. To simplify these calculations, we computed an effective distribution of transition rates corresponding to these sums.

The exponential of a normally-distributed random variable is log-normally distributed. Unfortunately, there is no analytical solution to a sum of log-normally distributed variables. One common approximation is the Fenton-Wilkinson approximation \cite{fenton1960sum}, which assumes that the distribution of a sum of log-normally distributed independent random variables is log-normally distributed, with a mean and variance equal to the sum of the mean and variance of the summed variables. This does not apply in our case as our distributions are truncated at a zero energy barrier, but we can apply the same idea. Consider a (truncated) normal distribution of reduced energy barriers (Eq.~\ref{eq:p0}),
\begin{equation}
P_1(U,\mu,\sigma) =  
\sqrt{\frac{2}{\pi}} \frac{e^{-\frac{(U-\mu)^2}{2 \sigma^2}}}{\sigma(1 + \text{erf}\left[ \frac{\mu}{\sqrt{2} \sigma} \right])},
\end{equation}
where $U>0$.
We can compute a distribution of rates $\rho_1(r,\mu,\sigma)$ for $r = e^{-U/k_B T}$:
\begin{align}
\begin{split}
    \rho_1(r,\mu,\sigma) & = -\frac{\partial}{\partial r}\left(\int_{0}^{-k_B T \log r} P_1(U,\mu,\sigma) dU\right)  \\
    & = k_B T \sqrt{\frac{2}{\pi}} \frac{e^{-\frac{(\mu + k_B T \log r)^2}{2 \sigma^2}}}{r \sigma(1 + \text{erf}\left[ \frac{\mu}{\sqrt{2} \sigma} \right])},
    \end{split}
\end{align}
noting that this distribution ranges from 0 to 1, as $U$ ranges from $0$ to $\infty$. 
From this distribution, we can compute the mean $m_1$, variance $v_1$, and third central moment $k_1$ of the rate distribution:
\begin{align}
    m_1 & = \int_0^1 r \rho_1(r,\sigma,\mu)dr =
    \frac{\text{erf}\left[\frac{\mu k_B T-\sigma ^2}{\sqrt{2} \sigma
   k_B T}\right]+1}{\text{erf}\left[\frac{\mu }{\sqrt{2} \sigma }\right]+1} e^{\frac{\sigma ^2-2 \mu  k_B T}{2 k_B^2 T^2}}\\
    v_1 & = \int_0^{1} (r-m_1)^2 \rho_1(r,\mu,\sigma)dr  = \frac{\text{erf}\left[\frac{\mu  k_BT-2 \sigma ^2}{\sqrt{2} \sigma 
   k_BT}\right]+1}{\text{erf}\left[\frac{\mu }{\sqrt{2} \sigma }\right]+1}e^{\frac{2 \sigma ^2-2 \mu  k_BT}{k_B^2 T^2}} - m_1^2 \\
   k_1 & = \int_0^{1} (r-m_1)^3 \rho_1(r,\mu,\sigma)dr = \frac {\text {erf}\left [\frac {k_B T \mu - 
          3 \sigma^2} {\sqrt {2} k_B T \sigma} \right] + 
     1 } {\text {erf}\left [\frac {\mu} {\sqrt {2} \sigma} \right] + 1} e^{\frac {9 \sigma^2 - 6 k_B T \mu} {2 k_B^2 T^2}} - m_1 \left( m^2_1 + 3 v_1 \right)
\end{align}
Direct measurement shows that the sum over $n$ rates resembles a lognormal distribution within the range $(0,1)$, but then rapidly decays for $r>1$. To model this, we use the piecewise function:
\begin{equation}
\rho_z(r,\sigma_z,\mu_z,C) = \frac{1}{N} \begin{cases}
 \frac{e^{-\frac{\left(k_BT \log (r)+\mu _z\right){}^2}{2 \sigma _z^2}}}{r} & 0\leq r\leq 1 \\
 e^{-\frac{\mu_z^2}{2 \sigma_z^2}} (1-\tanh (A (r-1)))
\end{cases}
\end{equation}
which, by definition, is continuous over the range $0$ to $\infty$ and converges to $0$ at large $r$. $N$ is a normalization constant and $A$ gives the rate of decay for the tail. Note that in reality the sum of $z$ rates should be bound to the range $0$ to $z$, but the difference is negligible because $r>z$ is extremely rare. $\mu_z$ and $\sigma_z$ can be thought of as effective $\mu$ and $\sigma$ values for an effective barrier that must be computed. To do this, we compute the mean $m_z$, variance $v_z$, and third central moment $k_z$ of this modified distribution. These integrals can be solved analytically, though the resulting expressions are cumbersome. 

\begin{figure}[t!]
\centering
\begin{tabular}[b]{c}
    \includegraphics[width=.45\linewidth]{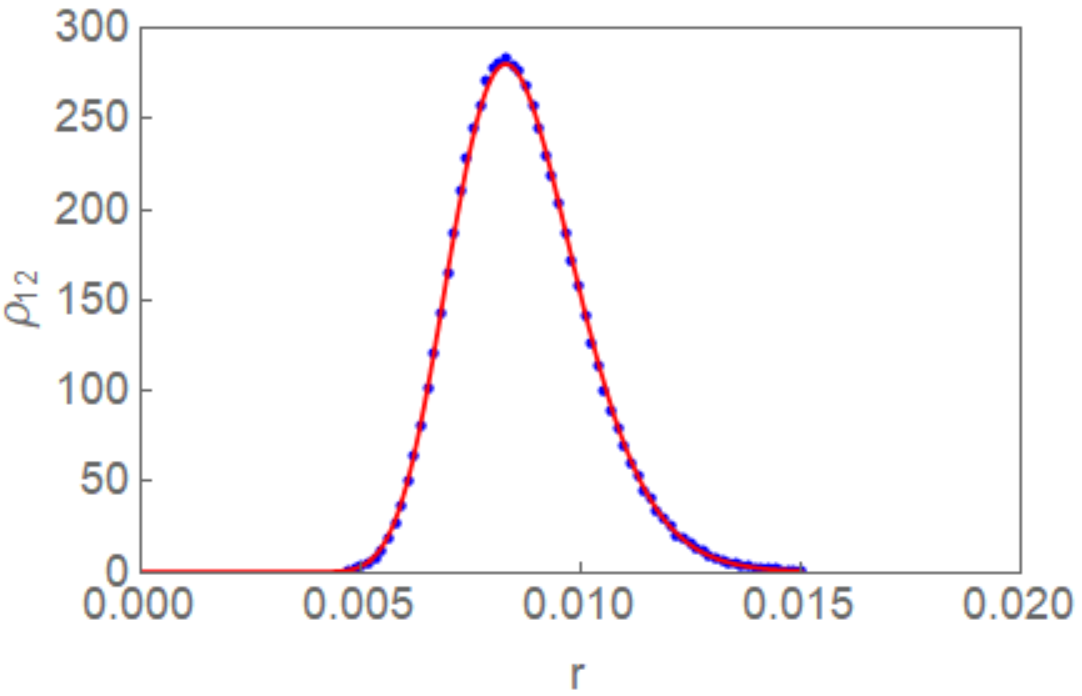} \\
    \small (a) $\rho_{12}$, $0.06$ eV
  \end{tabular} \quad
\begin{tabular}[b]{c}
    \includegraphics[width=.45\linewidth]{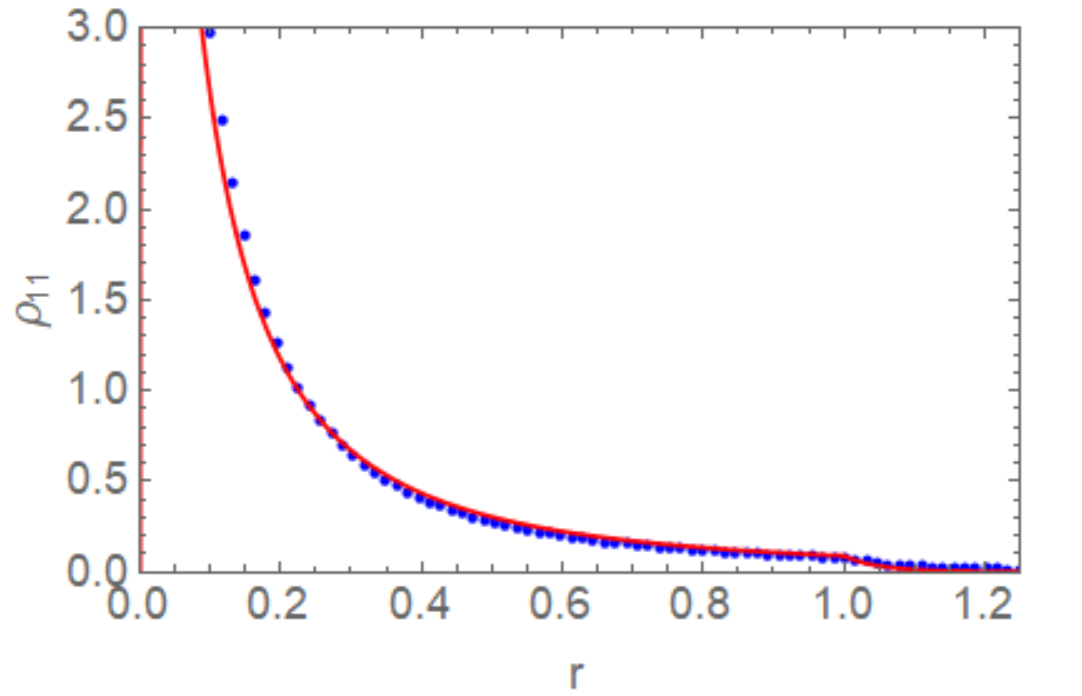} 
    \\
    \small (b) $\rho_{11}$, $0.32$ eV
    
  \end{tabular} 
\caption{Selected $\rho_{z}$ distributions from theory (blue) and computed numerically (red) for a) 12 rates at $\sigma=0.06$ eV and $\mu = 0.81$ eV and b) 11 rates at $\sigma=0.32$ eV and $\mu = 0.81$ eV.}
\label{fig:p10}
\end{figure}

To compute $\mu_z$, $\sigma_z$, and $A$, we use the fact that $m_z = z m_1$, $v_z = z v_1$, and $k_z = z k_1$. These simultaneous expressions can be solved numerically for given values of $\mu$, $\sigma$, $A$, and z without need for curve fitting or any prior simulation. $\rho_z$ distributions, compared with numerical simulation (computed by randomly generating sets of $z$ energies $U$ according to $P_1$ distributions and adding the corresponding rates) are given in Fig.~\ref{fig:p10}. Distributions like these were used for the effective rate terms in the primary text. Alternatively, one could numerically generate these rate sums and use fits for practical use, but this method is more readily applicable. 

\section{Higher-order correlation factors}
\label{sec:fcorr}

\begin{figure}[h!]
\centering
    \includegraphics[width=.65\linewidth]{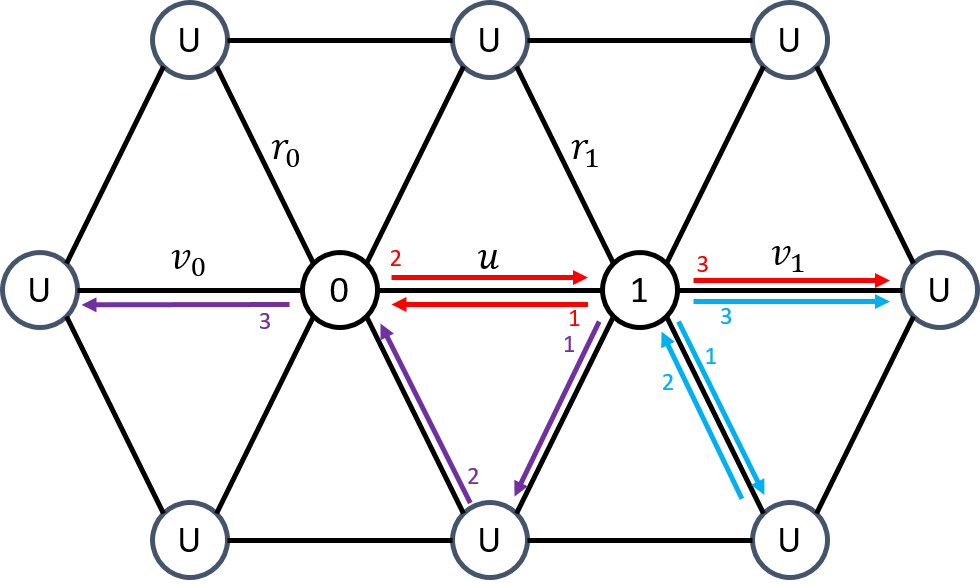}
\caption{Schematic vacancy migration environment. At time $i$, the vacancy occupies site $1$, having just hopped over a barrier $u$ from site $0$. All sites labeled U have barriers uncorrelated with the barriers around sites $1$ and $0$; note that some U sites are common neighbors of 0 and 1. The arrows represent three (red, blue, purple) possible trajectories contributing to the 3rd correlation factor $f_3$; the numbers at the base of the arrows indicates the order of hops. $r_0$ and $r_1$ represent effective rates as per the previous section. In the notation used in this section, red, blue, and purple would be $0\rightarrow101\text{U}$, $0\rightarrow 1\text{U}1\text{U}$, and $0\rightarrow1\text{U}0\text{U}$, respectively.}
\label{fig:traj}
\end{figure}

The correlation between subsequent hops is an essential factor in the computation of the diffusion constant. The introduction of random barriers introduces correlations because a vacancy that just hopped over a small barrier is more likely to hop back again over that same barrier. The $j^{\text{th}}$ correlation factor $f_j$ represents $\langle \vec{v}_i \cdot \vec{v}_{i+j} \rangle$, or the average cosine of the angle between a given hop and $j$ hops subsequent. The first correlation factor was given in Eq.~\ref{eq:f1} of the primary text:
\begin{equation}
    f_1 = \iiint \alpha_1(u,v,r)P_c(u,x)P_1(v,x)\rho_{z-2}(r,x) du dv dr,
\end{equation}
where
\begin{equation}
    \alpha_1 (u,v,r) = \frac{e^{-v}-e^{-u}}{e^{-v}+e^{-u}+r}.
\end{equation}
This can be interpreted as a Boltzmann sum with the $v$ term representing the chance that the vacancy hops in the same direction as the previous hop ( $\vec{v}_i \cdot \vec{v}_{i+1}=1$), the $u$ term representing the chance that the vacancy hops backwards over the previously-hopped barrier ( $\vec{v}_i \cdot \vec{v}_{i+1}=-1$), and the $r$ term representing the sum of the rates of other potential hops, which cancel in the numerator, but still contribute to the partition function (this fact, as well as $P_c$ being strongly biased toward small $u$ at large $x$, explains why the approximations for $\rho_{z-2}(r)$ are sufficient). For the other $\alpha$ terms, we make the following assumptions:
\begin{enumerate}
    \item A given hop, characterized by a barrier $u$, is only correlated with hops from the two sites that share that transition (i.e., the site occupied by the vacancy at time $i$ and the site occupied previously)
    \item Any hops that rearrange atomic species will randomly change the barriers, breaking the correlation 
\end{enumerate}
The above assumptions are true if the saddle points themselves are independent from each other(as in the KMC) and greatly reduce the number of trajectories we must consider. Figure~\ref{fig:traj} illustrates the environment around the vacancy and some of these trajectories. Any trajectories for which the last hop is from a U site do not contribute to the correlation factor, as assumption 1 dictates that all barriers around those sites are random and uncorrelated, meaning that the vacancy has an equal probability of hopping in any direction.

Given a set of barriers and rates $\{u,v_0,v_1,r_0,r_1\}$ corresponding to sites 0 and 1, one can compute $\alpha_j$ by adding the cosine between the previous hop and the final hop of all $j$-length trajectories (1 if parallel, -1 if anti-parallel, 0 by symmetry in all other cases), weighted by their probability. For example, 
\begin{equation}
\alpha_2(u,v_0,v_1,r_0,r_1) = \left(\frac{e^{-u}}{e^{-u}+e^{-v_1}+r_1} \right) \left(\frac{e^{-u}-e^{-v_0}}{e^{-u}+e^{-v_0}+r_0} \right),
\end{equation}
where the first term is the probability that the vacancy initially hops to site $0$ and the second is analogous to $\alpha_1$, but originating at site 0. The second correlation factor is then
\begin{equation}
f_2(x) = \int \dots \int \alpha_2 P_c(u)P_1(v_0)P_1(v_1)\rho_{z-2}(r_0)\rho_{z-2}(r_1)du dv_0 dv_1 dr_0 dr_1.
\end{equation}
The only set of trajectories that contribute to $f_2$ are the ones that start with a hop from 1 to 0. 
We will represent this by the notation $0\rightarrow10$, reflecting the previous site $0$, 
and all trajectories that begin with a hop from 1 to 0. While trivial now, it will be useful 
for the more complex trajectories needed for higher order factors. For example, a trajectory 
can include a hop from site 0 or 1 to a U site, followed by a hop immediately back ($0\rightarrow \text{U}0$ or $0\rightarrow10\text{U}0$) and still contribute to the correlation factor (e.g., the blue trajectory in Fig.~\ref{fig:traj}). A trajectory with a hop from 1 to U, then to 0 or vice-versa can contribute to the correlation factor if the U site is a common neighbor of 1 and 0 and both the U site and final destination site initially have the same atomic species ($1/m$ probability in an $m$-component equiatomic alloy, see the purple trajectory in Fig.~\ref{fig:traj}). If they share the same atomic species then after the two hops, the nearest neighbor environments are the same as a simple 01 hop.

\begin{figure}[h!]
\centering
    \includegraphics[width=.65\linewidth]{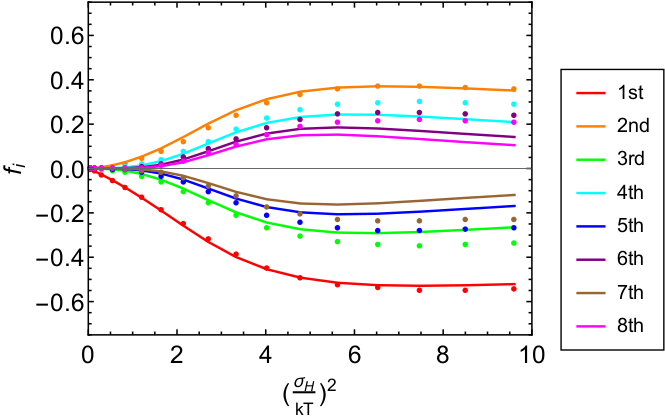}
\caption{Correlation factors computed from only the simplest trajectories, for which all but the last hops are between sites 0 and 1 (like the red trajectory in Fig.~\ref{fig:traj}).}
\label{fig:trajUncorr}
\end{figure} 

The simplest case for computing higher-order correlation factors is to consider only trajectories that involve direct back-and-forth hopping. In this case, odd-numbered factors end with a hop from site $1$ and even-numbered factors end with a hop from site $0$. In this case,
\begin{equation}
\alpha_i = \left(\frac{e^{-u}}{e^{-u}+e^{-v_1}+r_1}\right)^{\lfloor \frac{i}{2}\rfloor} \left(\frac{e^{-u}}{e^{-u}+e^{-v_0}+r_0} \right)^{\lceil \frac{i-2}{2}\rceil}\beta_i,
\end{equation}
where

\begin{equation}
    \beta_i = \begin{cases} \left( e^{-u}-e^{-v_0} \right) / \left( e^{-u}+e^{-v_0}+r_0 \right), \qquad i \text{ even} \\
    \left( e^{-v_1}-e^{-u} \right) / \left( e^{-u}+e^{-v_1}+r_1 \right), \qquad i \text{ odd}
    \end{cases}
\end{equation}

$\lceil \rceil$ and $\lfloor \rfloor$ represent the ceiling and floor functions, respectively. $r_1$ is the sum of all transition rates from site 1 not explicitly given, while $r_0$ is the same for site 0. The first factor arises from every hop from 1 to 0, and the second is for every hop from 0 to 1. $\beta_i$ is a reflection of whether the final hop is from site 0 or site 1. The result of calculations made with this $\alpha_i$ expression is given in Fig.~\ref{fig:trajUncorr} (which is exact for $f_1$ and $f_2$). This is a decent approximation for low-order factors or for small $\sigma$. However, as the CoNiFeCrMn MPEA exhibits large $\sigma_H$, it is insufficient for the alloy of interest. We must consider more complex trajectories that include U sites.

As the contributions of each trajectory to the correlation factor are additive, we can separate the integrals and treat different cases individually. At this point, it is worth considering the probability $p_{0\text{U}0}$ that the previous two hops were a 0U0 or 1U1 compound hop:
\begin{equation}
    p_{0\text{U}0}=p_{1\text{U}1}=11\frac{\int\dots\int \left( \frac{e^{-v}}{e^{-v}+e^{-u}+r_{z-2}} \right)^2 \left( \frac{e^{-v}}{e^{-v}+r_{z-1}} \right) P_c(u) P_1(v) \rho_{z-2}(r_{z-2}) \rho_{z-1}(r_{z-1})}{\int\dots\int \frac{e^{-v}}{e^{-u}+e^{-v}+r_{z-2}} P_c(u) P_1(v) \rho_{z-2}(r_{z-2})}. \label{eq:p0u0}
\end{equation}
$u$ is, again, the 01 barrier; in this case, $v$ represents the particular barrier between 0 or 1 and the site to which the vacancy bounces, while $r_{10}$ is the effective rates of all other sites from 0 or 1 and $r_{11}$ is the effective barrier of all other sites from U. The prefactor $11$ arises simply because there are $11$ U sites neighboring 1 or 0 in an FCC system. The square in the first integrand factor reflects the fact that among the 11 other neighbors, $v$ is more likely to be a small barrier and is more likely to result in a hop back to $1$ or $0$. This biased distribution is analogous to the $P_u$ distribution, except without an infinite sum it lacks an analytical normalization, hence the normalization term in the denominator. 1U0/0U1 hops do not require this biased distribution, but are only possible via the 4 U sites common to both 1 and 0. They also only contribute to the correlation factor if the U site and the final destination site have the same atomic species ($1/m$ chance for an $m$-component equiatomic alloy): 
\begin{equation}
    p_{0\text{U}1}=p_{1\text{U}0}=\frac{4}{5} \int\dots\int \left( \frac{e^{-v}}{e^{-v}+e^{-u}+r_{10}} \right) \left( \frac{e^{-v}}{e^{-v}+r_{11}} \right) P_u(u) P_1(v) \rho_{10}(r_{10}) \rho_{11}(r_{11}). \label{eq:p0u1}
\end{equation}

\begin{figure}[h!]
\centering
    \includegraphics[width=.65\linewidth]{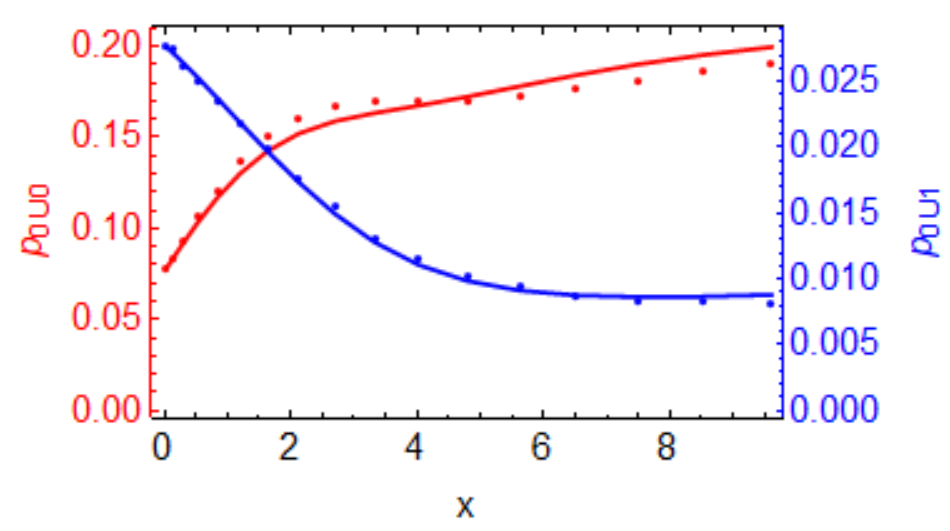} 
\caption{Average chance of a 1U1/0U0 (red) or 1U0/0U1 hop (blue) if the vacancy is initially at site 1/0, from KMC, compared with Eqs.~\ref{eq:p0u0} and \ref{eq:p0u1}.}
\label{fig:p0u0}
\end{figure}

\begin{figure}[h!]
\centering
    \includegraphics[width=.65\linewidth]{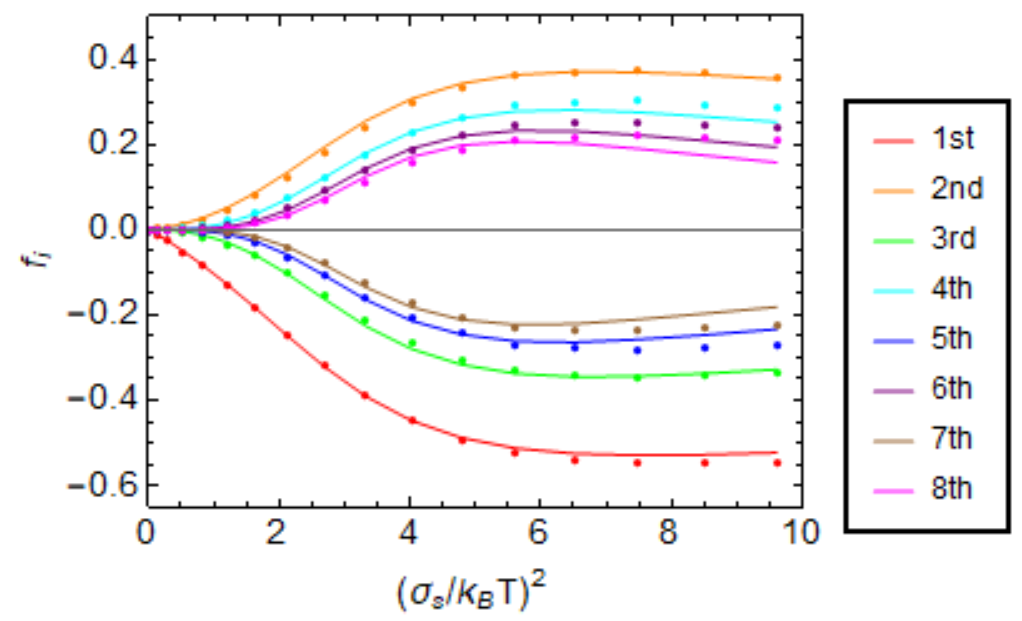}
\caption{Corrected correlation factors that include hops between sites 0 and 1 and other neighbors (like the blue and purple trajectories in Fig.~\ref{fig:traj}).}
\label{fig:trajCorr}
\end{figure}

Equations \ref{eq:p0u0} and \ref{eq:p0u1} are compared with KMC results in Fig.~\ref{fig:p0u0}.
For large $x$, $p_{0U0}$ approaches nearly $20\%$ when $\sigma = 0.34$ eV. For comparison, a 101 hop has a $\sim 60\%$ chance, meaning that when $\sigma$ is large, the vacancy has a nearly $80\%$ chance of revisiting the site occupied 2 hops prior. This extremely high level of correlation between hops necessitates this higher-order analysis.
We can use these terms to make a higher-order correction to $f_i$ with some final approximations. We will only consider trajectories that contain a single 0U0, 1U1, 0U1, or 1U0 hop (as $p_{0\text{U}0}^2 \approx 4\%$ at most), and we will assume that $p_{0U0}$, $p_{1U1}$, and $p_{0U1}$ is uncorrelated with $p_{01}$ (in truth, they should be weakly correlated). If this assumption holds, we can use the averaged $p_{0\text{U}0}$ and $p_{1\text{U}0}$ computed above. With these assumptions in place, what is the effect of a 1U1, 0U0, or 1U0 hop? A trajectory with a $1U1$ or $0U0$ hop is identical to a trajectory without, but with 2 fewer hops. A $1U0$ or $0U1$ hop is the same, except the penultimate site flips between 0 and 1 (which flips the $u$ barrier in Fig.~\ref{fig:traj} from left to right of the penultimate site, effectively flipping the sign of $\alpha_i$). The modified correlation factor for $i>2$ is then \begin{equation}
    f_i(x) = f^*_i(x) + (i-2)(p_{0U0}-p_{1U0})f^*_{i-2}(x), \label{eq:fFinal}
\end{equation}
where $f^*_i(x)$ is the correlation factor computed only from hops between sites 1 and 0 (Fig.~\ref{fig:trajUncorr}). The $(i-2)$ term is simply a count of the number of ways to arrange $i-2$ hops between 0 and 1 with one 0U0/1U1/0U1/1U0 hop. The corrected results given by Eq.~\ref{eq:fFinal} are given by Fig.~6b of the primary text and Fig.~\ref{fig:trajCorr} here. 

\end{document}